\documentclass[preprint]{aastex}

\begin{document}

\title{Observational Implications of a Plerionic Environment for Gamma-Ray Bursts}

\author{Dafne Guetta\altaffilmark{1} and Jonathan Granot\altaffilmark{2}}

\altaffiltext{1}{Osservatorio astrofisico di Arcetri, L.E. Fermi 2, Firenze, Italy; dafne@arcetri.astro.it}
\altaffiltext{2}{Institute for Advanced Study, Olden Lane, Princeton, NJ 08540; granot@ias.edu}

\begin{abstract}

We consider the possibility that at least some GRB explosions take place 
inside pulsar wind bubbles (PWBs), in the context of the supranova model,
where initially a supernova explosion takes place, leaving behind a supra-massive 
neutron star (SMNS), which loses its rotational energy on a time scale of months 
to tens of years and collapses to a black hole, triggering a GRB explosion. The 
most natural mechanism by which the SMNS can lose its rotational energy is 
through a strong pulsar type wind, between the supernova and the GRB events, 
which is expected to create a PWB. We analyze in some detail the observational
implications of such a plerionic environment on the afterglow and prompt GRB 
emissions, as well as the prospect for direct detection of the plerion emission. 
We find that for a simple spherical model, GRBs with iron lines detected in 
their X-ray afterglow should not have a detectable radio afterglow,
and should have small jet break times and non-relativistic transition times,
in disagreement with observations for some of the GRBs with X-ray lines. 
These discrepancies with the observations may be reconciled by resorting to a 
non-spherical geometry, where the PWB is elongated along the polar axis. 
We find that the emission from the PWB should persist well into the afterglow, 
and the lack of detection of such a component provides interesting constraints on 
the model parameters. Finally, we predict that the inverse Compton upscattering of
the PWB photons by the relativistic electrons of the afterglow (external Compton, EC) should 
lead to high energy emission during the early afterglow that might explain the GeV photons 
detected by EGRET for a few GRBs, and should be detectable by future missions such as GLAST. 

\end{abstract}


\section{Introduction}
\label{sec:intro}

The leading models for Gamma-Ray Bursts (GRBs) involve a relativistic wind emanating from a 
compact central source. The prompt gamma-ray emission is usually attributed to energy 
dissipation within the outflow itself, due to internal shocks within the flow that arise from 
variability in its Lorentz factor, while the afterglow emission arises from an external shock 
that is driven into the ambient medium, as it decelerates the ejected matter 
(Rees \& M\'esz\'aros 1994; Sari \& Piran 1997). In this so called `internal-external' shock model, 
the duration of the prompt GRB is directly related to the time during which the central source is 
active. The most popular emission mechanism is synchrotron radiation
from relativistic electrons accelerated in the shocks, that radiate in the strong magnetic fields
(close to equipartition values) within the shocked plasma. An additional radiation mechanism that 
may also play some role is synchrotron self-Compton (SSC), which is the upscattering 
of the synchrotron photons by the relativistic electrons, to much higher energies.

Progenitor models of GRBs are divided into two main categories. The first category involves the 
merger of a binary system of compact objects, such as a double neutron star (NS-NS, Eichler et al. 
1989), a neutron star and a black hole (NS-BH, Narayan, Pacy\'nski \& Piran 1992) or a black hole 
and a Helium star or a white dwarf (BH-He, BH-WD, Fryer \& Woosley 1998; Fryer, Woosley \& 
Hartmann 1999). The second category involves the death of a massive star. It includes the failed
supernova (Woosley 1993) or hypernova (Pacy\'nski 1998) models, where a black hole is created 
promptly, and a large accretion rate from a surrounding accretion disk (or torus) feeds a strong 
relativistic jet in the polar regions. This type of model is known as the collapsar model.
An alternative model within this second category is the supranova model (Vietri \& Stella 1998),
where a massive star explodes in a supernova and leaves behind a supra-massive neutron star (SMNS)
which on a time scale of a few years loses its rotational energy and collapses to a black hole, 
triggering the GRB event. Long GRBs (with a duration $\gtrsim 2\;{\rm s}$) are usually attributed 
to the second category of progenitors, while short GRBs are attributed to the first category. 
In all the different scenarios mentioned above, the final stage of the process consists of a newly
formed black hole with a large accretion rate from a surrounding torus, and involve a similar 
energy budget ($\lesssim 10^{54}\;{\rm ergs}$).

In this work we perform a detailed analysis of the supranova model, focusing on its possible 
observational signatures. This aims towards establishing tools that would enable us to 
distinguish between the supranova model and other progenitor models through observations,
and to constrain the model parameter using current observations.
The original motivation for the supranova model was to provide a relatively baryon clean 
environment for the GRB jet. As it turned out, it also seemed to naturally accommodate the 
later detection of iron lines in several X-ray afterglows (Lazzati, Campana, \& Ghisellini 1999; 
Piro et al. 2000; Vietri et al. 2001). 

It was later suggested that the most natural mechanism by which the SMNS can lose its 
rotational energy is through a strong pulsar type wind, between the supernova and the 
GRB events, which typically creates a pulsar wind bubble (PWB), also referred to as a 
plerion (K\"onigl \& Granot 2002, KG hereafter; Inoue, Guetta \& Pacini 2002, 
IGP hereafter). KG suggested that the shocked pulsar wind into which the afterglow 
shock propagates in this picture may naturally account for the large inferred values 
of $\epsilon_e\sim 0.1$ and $\epsilon_B\sim 10^{-3}-0.1$ (the  fractions of the internal 
energy in the electrons and in the magnetic field, respectively) that are inferred from 
fits to afterglow observations (Wijers \& Galama 1999; Granot, Piran \& Sari 1999; 
Chevalier \& Li 2000; Panaitescu \& Kumar 2002). This is attributed to the fact that 
pulsar winds are believed to largely consist of electron-positron pairs, and have
magnetization parameters in the right range. This relaxes the need of generating 
strong magnetic fields in the shock itself, as is required in other models, where 
the magnetic field in the external medium (assumed to be either the ISM or a stellar 
wind of a massive star progenitor) is typically too small to account for the values 
of $\epsilon_B$ that are inferred from observations. Another attractive feature of 
this model, pointed out by IGP is the possible high energy emission, in the GeV-TeV 
range, that may result from the upscattering of photons from the plerion by the 
relativistic electrons in the afterglow shock (external Compton, EC hereafter), 
and may be detected by GLAST. They have shown that the EC emission can provide a 
viable explanation for the extended GeV emission seen by EGRET in GRB 940217 
(Hurley et al. 1994).

We use a simple spherical model for the PWB. We find that a spherical model 
cannot accommodate the typical afterglow emission together with the iron line 
features observed in the X-ray afterglow of some GRBs. However, it was mentioned 
early on that in order to have a long lived afterglow emission together with the 
iron line features, a deviation from spherical symmetry is needed, where the line
of sight is relatively devoid of the material producing the iron lines 
(Lazzati et al. 1999; Vietri et al. 2001). This is required in order to avoid a direct 
collision of the afterglow shock with the line producing material on an observed 
time of the order of a day or so. It was later pointed out that a PWB is expected 
to exist inside the SNR shell, which decelerates the afterglow shock at a smaller 
radius, so that in order for the afterglow to remain relativistic up to a month 
or more, and produce the iron lines, we need the PWB to be elongated along its 
rotational axis (KG). In this paper we strengthen this conclusion, and show that 
in order to produce iron lines with a spherical PWB, its radius must be sufficiently 
small, resulting in a large density inside the PWB and a high self absorption 
frequency implying no radio afterglow, in contrast with observations. We leave the
detailed treatment of an elongated PWB to a future work, while in the present work 
we briefly comment about the expected effects of an elongated geometry compared
to a spherical one.

In this work we extend the analysis of KG and IGP, and perform detailed calculations of 
the radiation from the PWB, the prompt GRB and from the afterglow that occur inside the PWB.
We now give a short overview of the structure of the paper, where in each section we stress
the original features, new results and the observational constraints on the model. 
In \S \ref{PWB} we present our ``PWB'' model, introduce the relevant parameterization and 
model the acceleration of the supernova remnant (SNR) shell by the shocked pulsar wind. 
We use a simple spherical geometry and the pulsar wind is assumed to consist of proton 
and $e^\pm$ components with roughly equal 
energies, as well as a magnetic field. The conditions under which the iron line 
features that were observed in several X-ray afterglows may be reproduced within the PWB 
model, are investigated in \S \ref{lines}. We find that this requires a time delay of 
$\lesssim 1\;$yr between the supernova and the GRB events. In \S \ref{PlerEmis} we 
perform a detailed study of the plerion emission, including the synchrotron and SSC 
components, and provide an elaborate description of the relevant Klein-Nishina effect. 
We also discuss the upper cutoff that is imposed on high energy photons due to pair 
production with the radiation field of the PWB, go over the prospect for direct 
detection of the plerion emission, and derive observational constraints on the 
parameters of our model. The effects of the PWB environment on the prompt GRB 
emission are analyzed in \S \ref{prompt_GRB}, and we find that the EC from the 
prompt GRB should typically be very small, but might be detectable for extreme 
parameters. In \S \ref{Afterglow} we discuss the implications of a plerionic 
environment on the afterglow emission, and introduce the appropriate parameterization.
The radial density profile of the PWB is approximated as a power law in radius (KG), 
$\propto r^{-k}$, where $k$ typically ranges between $0$ (similar to an ISM) and $1$ 
(intermediate between an ISM and a stellar wind). The synchrotron, SSC and EC 
components are calculated and we provide detailed expressions for the break 
frequencies and flux normalization, for $k=0,\,1$. We also calculate the 
high energy emission that is predicted in this model. The results are discussed 
in \S \ref{discussion} and in \S \ref{conclusions} we give our conclusions.

\section{The Pulsar Wind Bubble}
\label{PWB}

Within the framework of the supranova model, a SMNS (also simply referred to as a pulsar) 
is formed in a supranova explosion, and then loses a large part of its 
rotational  energy before collapsing to a black hole and triggering the GRB. 
The most plausible mechanism for this energy loss is through a pulsar type wind 
(KG; IGP). A pulsar wind bubble (PWB) 
is formed when the relativistic wind (consisting of relativistic particles and 
magnetic fields) that emanates from a pulsar is abruptly decelerated (typically, 
to a Newtonian velocity) in a strong relativistic shock, due to interaction 
with the ambient medium. When a bubble of this type expands inside a SNR, 
it gives rise to a plerionic SNR, of which the Crab and Vela remnants are 
prime examples. Motivated by previous works (Rees \& Gunn 1974; Kennel \& Coroniti 1984; 
Emmering \& Chevalier 1987; KG) we consider in detail a spherical model where the shocked 
pulsar wind remains largely confined within the SNR. In \S \ref{discussion} briefly discuss 
the possible consequences of some more complicated geometries. 

The wind luminosity may be estimated by the magnetic dipole formula (Pacini 1967),
\begin{equation}\label{L_w}
L_w = {B_*^2R_*^6\Omega_*^4\over 6 c^3} = 7.0\times
10^{44} \left ({B_* \over 10^{12}\ G}\right )^2 \left({R_* \over
15\, {\rm km}}\right )^6 \left ({\Omega_* \over 10^4 \ {\rm
s}^{-1}} \right )^4\  {\rm ergs\ s^{-1}}\, ,
\end{equation}
where $B_*$ is the polar surface magnetic field, $R_*$ is the
circumferential radius (neglecting the distinction between its
equatorial and polar values in this approximation), and $\Omega_*$ is
the (uniform) angular velocity (whose maximum value is
$\sim 2\times 10^4\ {\rm s}^{-1}$; e.g., Haensel, Lasota, \& Zdunik 1999).
The spin-down time of a rapidly rotating SMNS can be estimated as
\begin{equation}\label{t_sd}
t_{\rm sd} = {E_{\rm rot} \over L_w} \approx 6 \left
( {\alpha \over 0.5}\right ) \left ({M_* \over 2.5 \, M_\odot}\right )^2
\left ( {R_* \over 15 \ {\rm km}} \right )^{-6} \left ({\Omega_*
\over 10^4\ {\rm s}^{-1}}\right )^{-3} \left ({B_* \over 10^{12}\
{\rm G}}\right )^{-2}\ {\rm yr}
\end{equation}
(see Vietri \& Stella 1998), where $E_{\rm rot} = \alpha G M_*^2 \Omega_*/2
c$ is the portion of the rotational energy of an SMNS of mass $M_*$ and angular velocity 
$\Omega_*$ that needs to be lost before it becomes unstable to
collapse.\footnote{The total rotational energy of the SMNS is given by 
$j G M_*^2 \Omega_*/2 c$, where the parameter $j$ measures
the stellar angular momentum in units of $GM_*^2/c$ and has values in the
range $0.57-0.78$ for realistic equations of state (e.g., Cook, Shapiro \& Teukolsky 1994; 
Salgado et al. 1994).} The spindown timescale, that sets the time delay between 
the supernova and GRB events, depends on the physical parameters of the SMNS.
Of these parameters, the least constrained is the magnetic field, which is typically
expected to be in the range of $\sim 10^{12}-10^{13}\;{\rm G}$, and may cause a variation 
of $\gtrsim 2$ orders of magnitude in $t_{\rm sd}$. There is also a strong dependence on the
radius $R_*$, which depends on its mass $M_*$ and the (uncertain) equation of state,
that may account for a change of up to $\sim 1$ order of magnitude in the scaling of
$t_{\rm sd}$. For example, for $R_*=10\;{\rm Km}$, with the values of the other parameters 
as given in Eq. (\ref{t_sd}), we have $t_{\rm sd}\approx 60\;{\rm yr}$. We conclude that the expected 
range of $t_{\rm sd}$ is from a few weeks to several years.

During $t_{\rm sd}$ the luminosity of the wind is roughly constant
and the wind should energize the PWB depositing an energy of the 
order of $E_{\rm rot}$. The luminosity of the pulsar wind is divided between its
different components: fractions $\xi_e$, $\xi_p$, and $\sigma_w$ in $e^\pm$ pairs, protons and
Poynting flux (magnetic field), respectively. The inferred values of $\xi_B=\sigma_w/(1+\sigma_w)$ 
for PWBs, such as Vela or the Crab, are typically $\sim 10^{-3}$ (Arons 2002), though there are 
also estimates as high as $\sim 1$ (Helfand, Gotthelf \& Halpern 2001). We shall adopt a fiducial 
value of $\sigma_w=10^{-3}$, which implies $\xi_e+\xi_p=1-\xi_B\cong 1$. Gallant \& Arons (1994) 
inferred $\xi_p/\xi_e\sim 2$ for the the Crab, and we adopt this estimate for our fiducial values,
and use $\xi_e=1/3$ and $\xi_p=2/3$. The inferred values of the Lorentz factor of the pulsar wind
are $\gamma_w\sim 10^4-10^7$.

The SN ejecta is accelerated by the force exerted due to the pressure of the expanding 
PWB, $p_{\rm out}$, at its outer boundary, i.e. at the radius of the SNR, $R_{\rm SNR}$. 
The pressure is expected to drop by a factor of order unity between the radius of the 
wind termination shock, $R_s$, and $R_{\rm SNR}$ (KG), so that 
$p_{\rm out}\equiv\eta_p p_{\rm av}$, where $p_{\rm av}=\eta E_{w}/3V=\eta E_{w}/4\pi R^3$ 
is the average pressure inside the PWB, $\eta$ is the fraction of the energy $E_{\rm rot}$ 
that remains in the PWB, $V=(4\pi/3)R^3$ is the volume of the PWB (which can be different 
for an elongated PWB) and $E_w\approx E_{\rm rot}(t/t_{\rm sd})$ is the energy emitted 
in the wind up to the time $t$. As we show below, the PWB is typically in fast cooling, 
and the electrons lose all of their internal energy to radiation, implying 
$\eta\approx\xi_p+\xi_B\approx\xi_p$. The equation of motion, 
$M_{\rm SNR}\ddot{R}=4\pi R^2p_{\rm out}$, may be written as
\begin{equation}\label{EOM}
R\ddot{R}={\eta_p\eta  E_{\rm rot}\over M_{\rm SNR}t_{\rm sd}}\,t\ .
\end{equation}

In the case of a non-spherical PWB, $M_{\rm SNR}$ should be replaced by the isotropic 
equivalent mass, $M_{\rm iso}(\theta)=4\pi dM_{\rm SNR}/d\Omega$, and $R$ becomes 
$R(\theta)$, where $\theta$ is the angle from the polar axis. However, the pressure 
can be taken as independent of the angle $\theta$, since the shocked pulsar wind is 
highly sub-sonic. If $M_{\rm iso}(\theta)$ is smaller near the poles ($\theta\approx 0$)
and larger near the equator ($\theta\approx\pi/2$), then even if the SNR shell is 
initially spherical, the acceleration would be larger near the poles, resulting in a 
much larger polar radius, $R_p$, compared to the equatorial radius, $R_{\rm eq}$: 
$R_p\gg R{\rm eq}$. This is a natural mechanism that can lead to an elongated geometry 
for the PWB.

Returning to the spherical case, the acceleration becomes significant at the time 
$t_{\rm acc}$ (and radius $R_{\rm acc}\approx v_0 t_{\rm acc}$)
when $E_w$ first exceed the initial kinetic energy of the SNR, 
$E_0=M_{\rm SNR} v_0^2/2$ ($\sim 10^{51}\,{\rm ergs}$), 
where $t_{\rm acc}=t_{\rm sd}E_0/E_{\rm rot}$.
We therefore have
\begin{equation}\label{ratios}
{t_{\rm sd}\over t_{\rm acc}}={E_{\rm rot}\over E_0}=
\left({v_b\over v_0}\right)^{2}=
\left({R_b\over R_{\rm acc}}\right)^{2/3}\approx 100\ ,
\end{equation}
where $v_b\equiv v_{\rm SNR}(t_{\rm sd})$ and $R_b\equiv R_{\rm SNR}(t_{\rm sd})$.
The dynamics are given by
\begin{equation}\label{SNR_dyn_R}
R_{\rm SNR}\approx\left\{\matrix{v_0 t & \ \ t<t_{\rm acc} \cr & \cr
R_b(t/t_{\rm sd})^{3/2} & \ \ t>t_{\rm acc} }\right. \ ,
\end{equation}
\begin{equation}\label{SNR_dyn_v}
v_{\rm SNR}\approx\left\{\matrix{v_0 & \ \ t<t_{\rm acc} \cr & \cr
v_b(t/t_{\rm sd})^{1/2} & \ \ t>t_{\rm acc} }\right. \ .
\end{equation}
These scalings agree with the results of Reynolds \& Chevalier (1984).
At $t>t_{\rm acc}$ we have $E_{\rm SNR}/E_w=3\eta_p\eta/2$, and conservation of energy 
implies that $E_0+E_w\approx E_w=\eta E_w +E_{\rm SNR}=(3\eta_p/2+1)\eta E_w$ or 
$(3\eta_p/2+1)\eta=1$. Our fiducial value of $\eta\approx\xi_p=2/3$ implies $\eta_p=1/3$,
which is reasonable. We also obtain that $E_{\rm SNR}(t_{\rm sd})=(3\eta/2)E_{\rm rot}\approx
E_{\rm rot}$. For a typical ejected mass, $M_{\rm SNR}\sim 10 M_{\odot}$, this would imply 
$v_b\sim 0.1c$. Finally we have
\begin{equation}
R_b = \frac{2}{3} v_b t_{\rm sd}= 6.3\times 10^{16} \beta_{b,-1} t_{\rm sd,0}
\end{equation}
where we set $v_b/c\equiv \beta_b =0.1 \beta_{b,-1}$ and $t_{\rm sd}=t_{\rm sd,0} \;{\rm yr}$.
To the extent that $v_b \propto (E_{\rm rot}/M_{\rm SNR})^{1/2}$ has nearly the same 
value in all sources, the magnitude of $R_b$ is determined by that of $t_{\rm sd}$. 
In a similar vien, if the energy lost during the SMNS lifetime, 
$E_{\rm rot}=10^{53}E_{53}\ {\rm ergs}$, is approximately constant from source 
to source ($E_{53} \sim 1$), then $t_{\rm sd}$ can also be used to
parameterize the SMNS wind power: $L_w = E_{\rm rot}/t_{\rm
sd}= 3.2 \times 10^{45}E_{53}/t_{\rm yr}\ {\rm ergs\
s^{-1}}$. 

The acceleration of the SNR shell by the lower-density bubble gas would subject 
it to a Rayleigh-Taylor (RT) instability, which could lead to clumping (Jun 1998). 
The growth timescale of the RT instability on a spatial scale R is 
$t_{\rm RT}\sim (R/\ddot{R})^{1/2}$. 
The important quantity to estimate in order to see if the SNR shell can be clumped
is the ratio between $t_{RT}$ and the dynamical timescale $t_{\rm dyn}=(R/\dot{R})$ 
\begin{equation}
\frac{t_{\rm RT}}{t_{\rm exp}}=\frac{\dot{R}}{(\ddot{R}R)^{1/2}}=
\sqrt{\frac{2 E_{\rm SNR}}{\eta_p\eta E_w}}\ ,
\end{equation}
where we have used Eq. (\ref{EOM}). This implies $t_{\rm RT}/t_{\rm exp}\approx\sqrt{3}$
during the acceleration ($t_{\rm acc}<t<t_{\rm sd}$).
This could produce only moderately strong fragmentation over the dynamical 
time of the system. However, as the acceleration occurs over $\sim 3$ orders 
of magnitude in radius [see Eq. (\ref{ratios})], the radius doubles itself $\sim 10$ 
times during the course of the SNR acceleration,
so that despite the fact that $t_{\rm RT}/t_{\rm exp}$ is of order unity, it is feasible that
considerable clumpiness may still be caused due to the RT instability. 
An even stronger fragmentation may occur if the RT instabilities 
grow on a length scale $x\sim \alpha R$ smaller than R (i.e. $\alpha<1$), where in 
this case $t_{\rm RT}/t_{\rm exp}\approx\sqrt{3\alpha}\,$.

In order to calculate the emission from the plerion, we use the average quantities of the
shocked pulsar wind within the PWB, and neglect their variation with radius. The latter 
is expected to be more important in the afterglow emission, and is therefore taken into account
in \S \ref{Afterglow} that discusses the afterglow emission.
The postshock energy density is given by
\begin{equation}\label{Edens}
e=\frac{\eta E_{\rm rot}}{V}
={3\eta E_{\rm rot}\over 4\pi R_b^3}\ ,
\end{equation}
where $V=(4\pi/3)R_b^3$ is the total volume within the PWB, and is approximately 
equal to the volume occupied by the shocked wind. For an elongated PWB the 
expression for the volume, $V$, will be different, but it can directly be plugged 
into these equations, in place of the spherical expression. The injection rates of 
electron-positron pairs and of protons at the source are given by
\begin{equation}\label{Ndot}
\dot{N}_{e,p}= {\xi_{e,p} L_w \over \gamma_w m_{e,p} c^2}\ .
\end{equation}
Hence, the total number of particles within $R_b$ at time $t$
is $N_{e,p}(t)=\dot{N}_{e,p}t$, and the number density at $t_{\rm sd}$ is
\begin{equation}\label{n_e,p}
n_{e,p}={N_{e,p}(t_{\rm sd})\over V}
={\xi_{e,p}\, e\over\eta\gamma_w m_{e,p}c^2}
={3\xi_{e,p} E_{\rm rot}\over 4\pi\gamma_w m_{e,p}c^2 R_b^3}\ .
\end{equation}
Fractions $\epsilon_{bB},\,\epsilon_{be},\,\epsilon_{bp}$ of the post-shock 
energy density go to the magnetic field, the electrons and the protons, respectively.
We expect these fractions to be similar to those in the pulsar wind ($\epsilon_{bB}\sim
\xi_B\sim\sigma_w$, $\epsilon_{be}\sim\xi_e$) and use the corresponding fiducial values. 
Subscripts containing the letter 'b' denote quantities related to the PWB.
The electrons will lose energy through synchrotron emission and 
inverse-Compton (IC) scattering. We study the characteristic features of the 
plerion emission and investigate the conditions required for the production of the 
observed iron lines and whether the plerion can be detected by the 
present instruments.   Moreover, an important implication of the plerion emission 
is that the GRB should explode inside a radiation rich environment
(i.e., the luminous radiation field of the PWB).
The external photons are highly Doppler-boosted in the rest frame of the shocked
fluid, for both internal and external shocks (that are responsible for the prompt 
GRB and afterglow emission, respectively), and can act as efficient seed
photons for IC scattering (external Comptonization, EC). We study the observational 
consequences of the EC process, both for the prompt GRB emission and for the afterglow.

Since we use a large number of parameters in the paper, and in order to make it 
easier to follow all the different parameters, we include a table (Table \ref{table1})
with the most often used parameters, where we mention the meaning of each parameter 
and the fiducial value that we use for that parameter.

\begin{deluxetable}{clc}
\tablecaption{The parameters most often used in the paper, their meaning and fiducial values.\label{table1}}
\tablehead{\colhead{Parameter} & 
\colhead{meaning} & \colhead{fiducial value}}
  \startdata
$t_{\rm sd}$ & time delay between SN and GRB & 1 or $10^{1.5}\;$yr \\
$\gamma_w$ & Lorentz factor of pulsar wind & $10^{4.5}$ \\ 
$E_{\rm rot}$ & rotational energy lost by SMNS & $10^{53}\;$erg \\
$M_{\rm SNR}$ & the mass of the SNR shell & $10\,M_\odot$ \\
$\eta$ & fraction of $E_{\rm rot}$ that remains in the PWB & $2/3$ \\
$\xi_B$ & fraction of wind energy in magnetic field & $10^{-3}$  \\ 
$\xi_e$ & fraction of wind energy in $e^\pm$ pairs & $1/3$ \\
$\xi_p$ & fraction of wind energy in protons  & $2/3$ \\
$\beta_b$ & velocity of SNR shell at $t_{\rm sd}$ (in units of $c$) & $0.1$ \\
$\epsilon_{bB}$ & fraction of PWB energy in magnetic field & $10^{-3}$ \\
$\epsilon_{be}$ & fraction of PWB energy in $e^\pm$ pairs & $1/3$ \\
$a$ & fraction of $e^\pm$ energy that is radiated & $1$ \\
$s$ & power-law index of $e^\pm$ distribution in PWB & $2.2$ \\
$p$ & electron power-law index in the GRB & $2.5$ \\
$\Gamma$ & bulk Lorentz factor of GRB ejecta & $10^{2.5}$ \\
$L_w$ & kinetic luminosity of GRB outflow & $10^{52}\;{\rm erg}/{\rm s}$ \\
$t_v$ & variability time of GRB central engine & $10\;$ms \\
$k$ & $n,\,e\propto r^{-k}$ in the PWB & 0 or 1 \\
$\epsilon_B$ & fraction of AG energy in magnetic field & $10^{-3}$ \\
$\epsilon_e$ & fraction of AG energy in electrons & $0.1$ \\
$E_{\rm iso}$ & isotropic equivalent energy in AG shock & $10^{53}\;$erg \\
$t$ & observed time since GRB & $1\;$day \\ \hline
\enddata
\end{deluxetable}

\section{X-ray Lines in the Afterglow}
\label{lines}
One of the main motivations for the supranova model is that it can naturally explain
the detections of iron lines in the X-ray afterglows of several GRBs, 
both in emission (GRB 970508, Piro et al. 1998; GRB 970828, Yoshida et al. 2001; GRB 000214, 
Antonelli et al. 2000; GRB 991216, Piro et al. 2000) and in absorption (GRB 990705; Amati et al. 
2000). The statistical significance of these detections is at the $\sim 3\sigma$ level, with the 
exception of GRB 991216 where a k-$\alpha$ emission line was detected with a significance of 
$\sim 4\sigma$. Emission lines of lighter elements (Mg, Si, S, Ar, Ca) have been reported in 
the X-ray afterglow of GRB 011211, at the level of $3\sigma$ (Reeves et al. 2002). This latter 
detection has been disputed by other authors (Borozdin \& Trudolyubov 2002; 
Rutledge \& Sako 2002), and may be said to be controversial. These line features may naturally 
arise in the context of the supranova model, where a SNR shell is located 
at a distance of $R\gtrsim 10^{16}\;{\rm cm}$ from the location of the GRB explosion
(Lazzati, Campana, \& Ghisellini 1999; Piro et al. 2000; Lazzati et al. 2001;
Vietri et al. 2001; B\"ottcher, Fryer, \& Dermer 2001). In this section we explore the 
condition under which such features may occur within our model, and obtain the relevant
constraints on the model parameters. In the following sections we investigate the 
implications of these constraints on the other observational signatures of the model: 
the plerion, prompt GRB and afterglow emissions.

We derive constraints on our model parameters using the observational data for
GRB 991216, as an example of an afterglow for which iron lines were detected,
since this is the most statistically significant detection to date.
Similar constraints may be obtained for other afterglows with X-ray features,
using similar arguments. The X-ray afterglow of GRB 991216 was observed by {\it Chandra}
from $37\;{\rm hr}$ to $40\;{\rm hr}$ after the GRB, and shows an emission 
line at $3.49\pm 0.06\;{\rm KeV}$, with a significance of $\sim 4\sigma$. The line flux 
was $F_{\rm Fe}\sim 1.6\times 10^{-13}\;{\rm erg\;cm^{-2}\;s^{-1}}$, which for a redshift
$z=1.02$ of this burst (Vreeswijk et al. 2000) implies an emission rate 
$\dot{N}_{\rm Fe}\sim 4\times 10^{52}\;s^{-1}$ of line photons, a luminosity of
$L_{\rm Fe}\sim 4\times 10^{44}{\rm erg\;s^{-1}}$ and a total energy of 
$E_{\rm Fe}\sim 3\times 10^{49}\;{\rm ergs}$ assuming the line emission lasted for 
$t_{\rm Fe}\sim 40/(1+z)\;{\rm hr}$ in the cosmological rest frame of the GRB 
(Vietri et al. 2001).

In the simplest version of our model, we assume a spherical geometry, and identify the 
line emitting material with the SNR shell, that is located at a radius $R_b$.
We use the above observations to derive constraints on $R_b$, or equivalently, on 
the time delay between the SN and the GRB events, $t_{\rm sd}$. The value of $R_b$ may 
be constrained by the requirement that the geometrical time delay in the arrival of the 
photons to the observer, $\sim R_b\theta_{\rm rad}^2/2$, should not exceed the total duration of 
the iron line emission, $t_{\rm Fe}$,
\begin{equation}\label{iron1}
R_b\lesssim 1.7\times 10^{18}\left({t_{\rm Fe}\over 20\,{\rm hr}}\right)
\left({\theta_{\rm rad}\over 0.05}\right)^{-2}\;{\rm cm}\quad,\quad 
t_{\rm sd}\lesssim 27\left({t_{\rm Fe}\over 20\,{\rm hr}}\right)
\left({\theta_{\rm rad}\over 0.05}\right)^{-2}\beta_{b,-1}^{-1}\;{\rm yr}\ ,
\end{equation}
where we have identified the opening angle to which the ionizing radiation extends, 
$\theta_{\rm rad}$, with the jet opening angle, $\theta_j\approx 0.05$ (Frail et al. 2001; 
Panaitescu \& Kumar 2002). 

Another constraint may arise from the requirement that $\dot{N}_{\rm Fe}=
N_{\rm Fe}/t_{\rm rec}\sim 4\times 10^{52}\;s^{-1}$, where the recombination time is 
given by $t_{\rm rec}\approx 4\times 10^9Z^{-2}T_e^{0.6}n_e^{-1}=2.8\times 10^{10}T_6^{0.6}
n_e^{-1}\;{\rm s}$, $T_6=T_e/10^6\,$K, and we have assumed an electric charge of 
$Z=24$ for the iron ions (Lazzati et al. 2001; KG).
In order to parameterize the electron number density, $n_e$, we need to relate between the 
width of the SNR shell, $\Delta R_b$, and its radius, $R_b$. If the SNR shell is 
efficiently fragmented during its acceleration phase, due to the RT instability, then
one might expect dense clumps of size $l_{cl}\ll\Delta R_b$, spread over a radial interval 
of $\Delta R_b$, that cover a fraction of order unity of the total solid angle.
This amounts to an effective width for the SNR shell of $\Delta R_{eff}=l_{cl}\ll\Delta R_b$,
implying $n_e=M_{\rm SNR}/4\pi R_b^2\Delta R_{eff}m_p$ and
\begin{equation}\label{iron2}
R_b\lesssim 9\times 10^{16}M_{\rm SNR,1}^{1/3}M_{\rm Fe,-1}^{1/3}
\xi_{-3}^{-1/3}T_6^{-1/5} \;{\rm cm}\quad,\quad 
t_{\rm sd}\lesssim 1.4M_{\rm SNR,1}^{1/3}M_{\rm Fe,-1}^{1/3}
\xi_{-3}^{-1/3}T_6^{-1/5}\beta_{b,-1}^{-1}\;{\rm yr}\ ,
\end{equation}
where $\xi\equiv\Delta R_{eff}/R_b=10^{-3}\xi_{-3}$ and $M_{\rm Fe}=0.1M_{\rm Fe,-1}M_\odot$ 
is the mass of the iron in the SNR shell. The SNR shell is compressed during its acceleration, 
and may attain $\Delta R_b\lesssim 0.1R_b$, so that $\xi$ may be as low as $\sim10^{-3}$.

A final constraint may be derived by considerations related to the total energy budget.
The total energy in the line is $E_{\rm Fe}\equiv\varepsilon E_\gamma\sim 3\times 10^{49}\;{\rm ergs}$
where the efficiency $\varepsilon$ is the product of the ratio of energies in the ionizing X-ray 
continuum and the prompt gamma-ray emission, and the energy fraction of the X-ray continuum 
that goes into the line emission, and is expected to be $\lesssim 0.01$ (Ghisellini et al. 2002).
This implies $E_\gamma\gtrsim 3\times 10^{51}\;{\rm ergs}$ which is somewhat in excess of the 
value $E_\gamma\approx 7\times 10^{50}\;{\rm ergs}$ found by Frail et al. (2001).
If the optical depth of the iron atoms is $\tau_{\rm Fe}<1$, then the efficiency is further reduced
by a factor of $\tau_{\rm Fe}=M_{\rm Fe}\sigma_{\rm Fe}/4\pi R_b^2 56 m_p$, where $\sigma_{\rm Fe}
\approx 2.0\times 10^{-20}\;{\rm cm^2}$ (Krolik \& Kallman 1987). Therefore, we must have 
$\tau_{\rm Fe}\gtrsim 1$, i.e.
\begin{equation}\label{iron3}
R_b\lesssim 6\times 10^{16}M_{\rm Fe,-1}^{1/2}\;{\rm cm}\quad,\quad 
t_{\rm sd}\lesssim 0.9M_{\rm Fe,-1}^{1/2}\beta_{b,-1}^{-1}\;{\rm yr}\ ,
\end{equation}

We conclude that X-ray line features, such as the ones observed in the afterglow 
of several GRBs, may be accommodate in a spherical model only for 
$t_{\rm sd}\lesssim 1\;{\rm yr}$, $R_b\lesssim 10^{17}\;{\rm cm}$.  
This constraint can be relaxed for an elongated PWB. In this case the iron lines
may be produced by the material near the equator, which is at a much smaller radius
than the polar radius, enabling the afterglow shock, that propagates along the poles, 
to reach a considerably larger radius.

\section{The Plerion Emission}
\label{PlerEmis}

In this section we evaluate the luminosity and the spectrum emitted by
the plerion. We use the average values of the quantities within the PWB,
that were derived in \S \ref{PWB}. As we show below, the electrons are
in the fast cooling regime for relevant values of $t_{\rm sd}$, and therefore
most of the emission takes place within a small radial interval just behind 
the wind termination shock, and the various quantities should not vary 
significantly within this region, and should not be very different from their 
average values within the PWB.

The magnetic field inside the plerion is 
\begin{equation}\label{B_b}
B_b=\sqrt{8\pi \epsilon_{bB} e}\approx 1.3\,\eta_{2/3}^{1/2}\epsilon_{bB,-3}^{1/2}
E_{\rm rot,53}^{1/2}\beta_{b,-1}^{-3/2}t_{\rm sd,0}^{-3/2}\;{\rm G}\ .
\end{equation}
where $\epsilon_{bB,-3}=\epsilon_{bB}/10^{3}$ and $\eta_{2/3}=\eta/(2/3)$.
Relativistic electrons/positrons (hereafter simply electrons)
are injected into the plerion at the rate $\dot{N_e}$ (see Eq.\ref{Ndot})with a power law
distribution $N(\gamma_b)\equiv dn/d\gamma_b\propto \gamma_b^{-s}$ in the 
Lorentz factor (LF) range $\gamma_{bm}\lesssim \gamma_b\lesssim\gamma_{bM}.$
The minimum electron LF is given by  
\begin{equation}\label{gamma_bm}
\gamma_{bm}=\left({s-2\over s-1}\right){\epsilon_{be}e\over n_e m_e c^2}
\approx 3.5\times 10^3 \eta_{2/3}\epsilon_{be,1/3}\xi_{e,1/3}^{-1}\gamma_{w,4.5}\ ,
\end{equation}
where $\gamma_{w,4.5}=\gamma_{w}/10^{4.5}$, 
$\epsilon_{be,1/3}=\epsilon_{be}/(1/3)$, $\xi_{e,1/3}=\xi_{e}/(1/3)$ 
and we use $s=2.2$ to obtain the numerical values for the rest of the paper.
The electrons radiatively cool by the combination of the synchrotron
and synchrotron-self-Compton (SSC) process, the timescales of which are
$t_{\rm syn}\sim 6\pi m_e c/\sigma_T B_b^2 \gamma_b$  and 
$t_{SC}=t_{\rm syn}/Y_b$ and the combined cooling time being 
$t_c=(1/t_{\rm syn}+1/t_{SC})^{-1}=t_{\rm syn}/(1+Y_b)$, where 
\begin{equation}\label{Y_b}
Y_b\sim{a\epsilon_{be}\over(1+Y_b)\epsilon_{bB}}
\quad,\quad Y_b\approx\left\{\matrix{a\epsilon_{be}/\epsilon_{bB} & 
a\epsilon_{be}/\epsilon_{bB}\ll 1 \cr & \cr
\sqrt{a\epsilon_{be}/\epsilon_{bB}} & a\epsilon_{be}/\epsilon_{bB}\gg 1}\right. \ ,
\end{equation}
(Sari, Narayan \& Piran 1996; Panaitescu \& Kumar 2000; Sari \& Esin 2001)\footnote{
Sari and Esin pointed out that generally the factor $a$ should be multiplied by the 
fluid velocity just behind the shock (in the shock frame), which for a relativistic 
shock with $\epsilon_B\ll 1$, that is relevant for the pulsar wind termination shock, 
is $\beta=1/3$. This factor of $1/3$ should be divided by the ratio of the radiation 
flux and the photon energy density times $c$, which is $1/4$ for an isotropic emission 
in the local rest frame of the emitting fluid. Together this gives a factor $4/3$ which 
is close to 1, and is therefore neglected.}
is the Compton y-parameter of the plerion, which is the fractional energy gain of a 
photon when traveling through the plerion, due to upscattering by the relativistic 
electrons, and $a\equiv\min[1,(\gamma_{bm}/\gamma_{bc})^{s-2}]$ is the fraction of 
the internal energy in the electrons that is radiated away (Sari \& Esin 2001). 
For our choice of parameters $\epsilon_{bB}\ll\epsilon_{be}$ and there is fast cooling 
so that $a=1$. This implies $Y_b\approx(\epsilon_{be}/\epsilon_{bB})^{1/2}$, and we shall 
use this relation in the following. 
The maximum LF is set by equating $t_c$ with the acceleration time, 
$\sim 2\pi \gamma_b m_e c/q B_b$ (where $q$ is the electric charge of 
the electron):
\begin{equation}\label{gamma_bM}
\gamma_{bM}=\sqrt{3 q\over B_b \sigma_T(1+Y_b)}\approx 
9.7\times 10^6 \,a^{-1/4}\eta_{2/3}^{-1/4}\epsilon_{be,1/3}^{-1/4} 
E_{\rm rot,53}^{-1/4}\beta_{b,-1}^{3/4}t_{\rm sd,0}^{3/4}\ .
\end{equation}
The LF of an electron that cools on the adiabatic expansion time $t_{\rm sd}$ 
(i.e. an electron for which $t_c=t_{\rm sd}$) is given by
\begin{equation}\label{gamma_bc}
\gamma_{bc}={6\pi m_e c\over (1+Y_b)B_b^2 \sigma_T t_{\rm sd}}\approx
0.84\,a^{-1/2}\eta_{2/3}^{-1}\epsilon_{be,1/3}^{-1/2} \epsilon_{bB,-3}^{-1/2}E_{\rm rot,53}^{-1}
\beta_{b,-1}^{3}t_{\rm sd,0}^{2}\ .
\end{equation}
For $t_{\rm sd}<t_1$, where
\begin{equation}\label{t_1}
t_1\approx 1.1\,\eta_{2/3}^{1/2}\epsilon_{be,1/3}^{1/4}\epsilon_{bB,-3}^{1/4}
E_{\rm rot,53}^{1/2}\beta_{b,-1}^{-3/2} \;{\rm yr}\ ,
\end{equation}
the cooling is so fast that Eq. (\ref{gamma_bc}) implies $\gamma_{bc}<1$. 
This means that all electrons cool to non-relativistic random velocities within a 
time $\sim\gamma_{bc}t_{\rm sd}$ and a distance $\sim\gamma_{bc}R_b$ behind the 
termination shock of the pulsar wind. Of course, in this regime $\gamma_{bc}<1$ 
no longer corresponds to the physical LF of the electrons\footnote{we formally 
obtain $\gamma_{bc}<1$ because we used the approximation $\beta^2\approx 1$ in the 
expression for the total synchrotron power of a single electron, under the 
assumption (that does not hold here) that the electrons are always relativistic.}. 
We shall call this regime very fast cooling.

In the case of a non-spherical (elongated) PWB, we can still use the same expressions,
if we make the simple substitution described below. We define an effective radius,
$R_{\rm eff}\equiv(3V/4\pi)^{1/3}$, so that a sphere of that radius will have the same 
volume, $V$, as the non-spherical PWB. Since the volume of the PWB determines its
average number density and energy density (which determine the PWB emission), and 
since we expect the interior of the PWB 
to be roughly homogeneous (i.e. the local values are not very different from the mean 
values), then we can reproduce the expressions appropriate for a non-spherical model
by simply replacing $R_b$ with $R_{\rm eff}$. Since $\beta_b$ appears in the various 
expressions only through $R_b$, we may achieve this task most easily by substituting 
$\beta_b=3R_{\rm eff}/2ct_{\rm sd}$ everywhere. In order to illustrate the effects of
a non-spherical geometry, we consider a particular example (which we also consider to 
be likely) of an elongated PWB with a polar radius much larger than the equatorial radius, 
$R_p\gg R_{\rm eq}$, where $R_p/2\lesssim R_{\rm eff}< R_p$. If the surface mass density 
at the poles is sufficiently small, then the velocity of the SNR shell there can become 
close to $c$, so that $R_p\approx c t_{\rm sd}$. This would imply
$\beta_b=3R_{\rm eff}/2ct_{\rm sd}\approx 1$

\subsection{The Synchrotron Spectrum}
\label{plerion_syn}

The characteristic synchrotron frequency of an electron with LF $\gamma_b$ is 
$\nu_b=\gamma_b^2 \nu_{b0}$ where $\nu_{b0}=3 q B_b/16 m_e c$. The synchrotron frequencies 
corresponding to $\gamma_{bm},\gamma_{bM}$ and $\gamma_{bc}$, respectively, are
\begin{eqnarray} 
\nu_{bc}&\approx & 2.9\times 10^{6}\,(1+z)^{-1}a^{-1}\eta_{2/3}^{-3/2}\epsilon_{be,1/3}^{-1}
\epsilon_{bB,-3}^{-1/2}E_{\rm rot,53}^{-3/2}\beta_{b,-1}^{9/2}t_{\rm sd,0}^{5/2}\;{\rm Hz}\ .
\nonumber \\
\nu_{bm}&\approx & 5.1\times 10^{13}\,(1+z)^{-1}\eta_{2/3}^{5/2}\epsilon_{be,1/3}^2
\epsilon_{bB,-3}^{1/2}\xi_{e,1/3}^{-2} E_{\rm rot,53}^{1/2}\gamma_{w,4.5}^2\beta_{b,-1}^{-3/2}
t_{\rm sd,0}^{-3/2}\;{\rm Hz}\ ,
\\
\nu_{bM}&\approx & 3.9\times 10^{20}\,(1+z)^{-1}a^{-1/2}
\epsilon_{be,1/3}^{-1/2}\epsilon_{bB,-3}^{-1/2}{\rm Hz}\ ,
\nonumber
\end{eqnarray}

For $t_{\rm sd}<t_2$ [which is given in Eq. (\ref{t_2})] the synchrotron self 
absorption frequency $\nu_{bsa}$ (which is treated below) is above the 
cooling frequency $\nu_{bc}$, and the synchrotron flux density, $F_{\nu}$, 
peaks at $\nu_{bsa}$ and consists of three power-law segments:
\begin{equation}\label{Fnu_syc_VFC}
{F_{\nu}(t_{\rm sd}<t_2)\over F_{\nu,{\rm max}}}=\left({\nu_{bsa}\over\nu_{bc}}\right)^{-1/2}
\times\left\{\matrix{
(\nu/\nu_{bsa})^{2} & \ \ \nu<\nu_{bsa} \cr & \cr 
(\nu/\nu_{bsa})^{-1/2} & \nu_{bsa}<\nu<\nu_{bm} \cr & \cr 
(\nu_{bm}/\nu_{bsa})^{-1/2}(\nu/\nu_{bm})^{-s/2} & \nu_{bm}<\nu<\nu_{bM}}\right. \ .
\end{equation}
For time separations between the SN and GRB events, $t_1<t_{\rm sd}<t_3$, where
\begin{equation}\label{t_3}
t_3\approx 65\,\eta_{2/3}\epsilon_{be,1/3}^{3/4}\epsilon_{bB,-3}^{3/4}\xi_{e,1/3}^{-1/2}
 E_{\rm rot,53}^{1/2}\gamma_{w,4.5}^{1/2}\beta_{b,-1}^{-3/2}\;{\rm yr}\ ,
\end{equation}
we have $1<\gamma_{bc}<\gamma_{bm}$ and the bubble is in the (moderately) 
fast cooling regime. For $t_2<t_{\rm sd}<t_3$, $F_{\nu}$
peaks at $\nu_c$ and is given by
\begin{equation}\label{Fnu_syc_FC}
{F_{\nu}(t_2<t_{\rm sd}<t_3)\over F_{\nu,{\rm max}}}=\left\{\matrix{
(\nu_{bsa}/\nu_{bc})^{1/3}(\nu/\nu_{bsa})^2 & \ \ \nu<\nu_{bsa} \cr & \cr 
(\nu/\nu_{bc})^{1/3} & \ \ \nu_{bsa}<\nu<\nu_{bc} \cr & \cr 
(\nu/\nu_{bc})^{-1/2} & \nu_{bc}<\nu<\nu_{bm} \cr & \cr 
(\nu_{bm}/\nu_{bc})^{-1/2}(\nu/\nu_{bm})^{-s/2} & \nu_{bm}<\nu<\nu_{bM}}\right. \ .
\end{equation}
For $t_{\rm sd}>t_3$ the bubble is in the slow cooling regime, where the 
spectrum peaks at $\nu_{bm}$ and again consists of four power law segments:
\begin{equation}\label{Fnu_syc_SC}
{F_{\nu}(t_{\rm sd}>t_3)\over F_{\nu,{\rm max}}}=\left\{\matrix{
(\nu_{bsa}/\nu_{bm})^{1/3}(\nu/\nu_{bsa})^2 & \ \ \nu<\nu_{bsa} \cr & \cr 
(\nu/\nu_{bm})^{1/3} & \ \ \nu_{bsa}<\nu<\nu_{bm} \cr & \cr 
(\nu/\nu_{bm})^{(1-s)/2} & \nu_{bm}<\nu<\nu_{bc} \cr & \cr 
(\nu_{bc}/\nu_{bm})^{(1-s)/2}(\nu/\nu_{bc})^{-s/2} & \nu_{bc}<\nu<\nu_{bM}}\right. \ .
\end{equation}
The peak synchrotron flux, $F_{\nu,{\rm max}}$, for either fast
or slow cooling, is 
\begin{equation}
F_{\nu,{\rm max}}\approx{\xi_e E_{\rm rot}P_{\nu,{\rm max}}\over\gamma_w m_e c^2}
{(1+z)\over 4\pi d_L^2}
\approx 42\,(1+z)\eta_{2/3}^{1/2}\epsilon_{bB,-3}^{1/2}\xi_{e,1/3}E_{\rm rot,53}^{3/2}
\beta_{b,-1}^{-3/2}t_{\rm sd,0}^{-3/2}\gamma_{w,4.5}^{-1}d_{L28}^{-2}\;{\rm mJy}\ , 
\end{equation}
where $P_{\nu,{\rm max}}\approx P_{e,{\rm syn}}/\nu_{\rm syn}$, 
$P_{e,{\rm syn}}=(4/3)\sigma_Tc(B_p^2/8\pi)\gamma_e^2$, $\nu_{\rm syn}=\nu_{b0}\gamma_e^2$ 
and $d_L=10^{28}d_{L28}\,{\rm cm}$ is the the luminosity distance of the GRB.
Synchrotron self-absorption (SSA) will cause a break in the spectrum at a frequency $\nu_{bsa}$ 
below which\footnote{for PWBs, this applies also for the fast cooling regime, as opposed to the 
spectral slope of $F_{\nu}\propto\nu^{11/8}$ (Granot, Piran \& Sari 2000; Granot \& Sari 2002) 
predicted for GRBs (both for the prompt emission and afterglow) since for the PWB the observer 
faces the back of the shell, i.e. the side farther from the shock, where the coolest electrons 
reside.} $F_{\nu}\propto\nu^2.$ The absorption coefficient is given by:
\begin{equation}
\alpha_{\nu}=-\frac{1}{8\pi m_e\nu^2}\int
d\gamma_e P_{\nu,e} \gamma_e^2 \frac{\partial}{\partial\gamma_e}
\left[\frac{N(\gamma_e)}{\gamma_e^2}\right]
\end{equation}
where 
\begin{equation}
P_{\nu,e}\approx P_{\nu,{\rm max}}\left(\frac{\nu}{\nu_{\rm syn}}\right)^{1/3}\approx
\frac{8m_e c^2 B_p\sigma_T}{9\pi q} \left(\frac{\nu}{\nu_{\rm syn}}\right)^{1/3}
\quad {\rm for}\quad (\nu<\nu_{\rm syn})
\end{equation}
is the spectral emissivity of a single electron.
 
The electron distribution, $N(\gamma_e)$, is different for the fast cooling
and slow cooling regimes. In the fast cooling case we have 
\begin{equation}\label{N(gam)_FC}
N(\gamma_e)\approx N_0\times
\left\{\matrix{\gamma_e^{-2} & \ \ \max(\gamma_{bc},1)<\gamma_e<\gamma_{bm} 
\cr & \cr
\gamma_{bm}^{-2}(\gamma_e/\gamma_{bm})^{-s-1} & \ \ 
\gamma_{bm}<\gamma_e<\gamma_{bM}
}\right. \ ,
\end{equation}
with $N_0=\gamma_{bc} n_e$. The local electron distribution depends on the 
distance behind the shock, $l$, and the distribution given above is averaged 
over $l$. Furthermore, it includes only the relativistic electrons (i.e. for 
$\gamma_c<1$, most electrons would cool to non-relativistic random Lorentz 
factors, and are not included in this distribution).

For $\nu<\max(\nu_{bc},\nu_{b0})$, all the electrons with $\gamma_e>
\max(\gamma_c,1)$ contribute to the SSA and the absorption coefficient is
\begin{equation}\label{alpha1}
\alpha_{\nu}\approx\frac{3 N_0 P_{\nu,{\rm max}}}{16\pi m_e\nu^{5/3}\nu_{b0}^{1/3}}
\min(\gamma_{bc}^{-8/3},1)\ .
\end{equation}
If $\max(\nu_{bc},\nu_{b0})<\nu<\nu_{bm}$ we have
\begin{equation}\label{alpha2}
\alpha_{\nu}\approx\frac{3 N_0 P_{\nu,{\rm max}}}{16\pi m_e\nu^{5/3}\nu_{b0}^{1/3}}
\left(\frac{16 m_e c\nu}{3 q B_b}\right)^{-4/3}\ ,
\end{equation}
and, finally, if $\nu>\nu_{bm}$ then
\begin{equation}\label{alpha3}
\alpha_{\nu}\approx
\frac{3 N_0 P_{\nu,{\rm max}}}{16\pi m_e\nu^{5/3}\nu_{b0}^{1/3}}
\left(\frac{16 m_e c\nu}{3 q B_b}\right)^{-(s+5/3)/2}\gamma_{bm}^{s-1}\ .
\end{equation}
In the slow cooling case we have 
\begin{equation}\label{N(gam)_SC}
N(\gamma_e)\approx N_0\times
\left\{\matrix{\gamma_e^{-s} & \ \ \gamma_{bm}<\gamma_e<\gamma_{bc} 
\cr & \cr
\gamma_{bc}^{-s}(\gamma_e/\gamma_{bc})^{-s-1} & \ \ 
\gamma_{bc}<\gamma_e<\gamma_{bM}
} \right. \ .
\end{equation}
with $N_0=(s-1)\gamma_{bm}^{s-1} n_e$.
The absorption coefficient for $\nu<\nu_{bm}$ is given by
\begin{equation}\label{alpha4}
\alpha_{\nu}=\frac{4}{(3s+2)}
\frac{N_0 P_{\nu,{\rm max}}}{8\pi m_e\nu^{5/3}\nu_{b0}^{1/3}}\gamma_{bm}^{-s-2/3}
\end{equation}
The optical depth to SSA is given by $\tau_{\nu,sa}=\alpha_{\nu}R_b\,{\rm min}
(\gamma_{bc},1)$, and $\nu_{sa}$ is obtained by equating $\tau_{\nu,sa}=1$.
For $t_{\rm sd}<t_2$, where 
\begin{equation}\label{t_2}
t_2\approx 12\,\eta_{2/3}^{3/7}\epsilon_{be,1/3}^{5/21}\epsilon_{bB,-3}^{4/21}
\xi_{e,1/3}^{2/21}E_{\rm rot,53}^{11/21}\gamma_{w,4.5}^{-2/21}\beta_{b,-1}^{-31/21}\;{\rm yr}\ ,
\end{equation}
we have $\nu_{bsa}>\nu_{bc}$. Since typically $t_1<t_2<t_3$, for $t_{\rm sd}<t_2$ 
we have either very fast cooling or moderately fast cooling and may use 
Eq. (\ref{alpha2}) to derive
\begin{equation}\label{nusa2}
\nu_{bsa}\approx 1.9\times 10^{10}\,(1+z)^{-1}a^{-1/6}\epsilon_{be,1/3}^{-1/6}\epsilon_{bB,-3}^{1/6}
\xi_{e,1/3}^{1/3}E_{\rm rot,53}^{1/3}\gamma_{w,4.5}^{-1/3}
\beta_{b,-1}^{-2/3}t_{\rm sd,0}^{-1}\;{\rm Hz}\ .
\end{equation}
For $t_2<t_{\rm sd}<t_3$ we use Eq. (\ref{alpha1}) and obtain
\begin{equation}\label{nusa1}
\nu_{bsa} \approx  1.1\times 10^{13}\,(1+z)^{-1}a^{1/2}\eta_{2/3}^{6/5}\epsilon_{be,1/3}^{1/2}
\epsilon_{bB,-3}^{7/10}\xi_{e,1/3}^{3/5}E_{\rm rot,53}^{9/5}\gamma_{w,4.5}^{-3/5}
\beta_{b,-1}^{-24/5}t_{\rm sd,0}^{-19/5}\;{\rm Hz}\ .
\end{equation}
Since the SSA frequency approaches $\nu_{bm}$ only for very large time delays 
$t_{\rm sd}\gg t_3$, we do not consider the case $\nu_{bsa}>\nu_{bm}$.
In the slow cooling regime ($t_{\rm sd}>t_3$) we have 
\begin{equation}\label{nusa3}
\nu_{bsa} \approx  2.9\times  10^{9}\,(1+z)^{-1}\eta_{2/3}^{-4/5}\epsilon_{bB,-3}^{1/5}
\xi_{e,1/3}^{8/5}E_{\rm rot,53}^{4/5}\gamma_{w,4.5}^{-8/5}
\beta_{b,-1}^{-9/5}t_{\rm sd,0}^{-9/5}\;{\rm Hz}\ .
\end{equation}

\subsection{The SSC Spectrum}
\label{plerion_SSC}

The SSC spectrum has a similar shape to the synchrotron spectrum, and 
we approximate it as comprising of broken power-laws with characteristic 
frequencies. For the different ranges in $t_{\rm sd}$, it assumes the 
following forms:
\begin{equation}\label{Fnu_syc_SC1}
{\nu F^{SC}_{\nu}(t_{\rm sd}<t_2)\over Y_b\,\nu_{bm} F_{\nu_{bm}}}=\left\{\matrix{
(\nu^{SC*}_{bsa}/\nu^{SC}_{bm})^{1/2}(\nu/\nu^{SC*}_{bsa})^2 & \ \ \nu<\nu^{SC*}_{bsa} \cr & \cr 
(\nu/\nu^{SC}_{bm})^{1/2} & \nu^{SC*}_{bsa}<\nu<\nu^{SC}_{bm} \cr & \cr 
(\nu/\nu^{SC}_{bm})^{(2-s)/2} & \nu>\nu^{SC}_{bm}}\right. \ .
\end{equation}
\begin{equation}\label{Fnu_syc_SC2}
{\nu F^{SC}_{\nu}(t_2<t_{\rm sd}<t_3)\over Y_b\,\nu_{bm} F_{\nu_{bm}}}=\left\{\matrix{
(\nu^{SC}_{bc}/\nu^{SC}_{bm})^{1/2}(\nu^{SC}_{bsa}/\nu^{SC}_{bc})^{4/3}(\nu/\nu^{SC}_{bsa})^2 & 
\ \ \nu<\nu^{SC}_{bsa} \cr & \cr 
(\nu^{SC}_{bc}/\nu^{SC}_{bm})^{1/2}(\nu/\nu^{SC}_{bc})^{4/3} & 
\nu^{SC}_{bsa}<\nu<\nu^{SC}_{bc} \cr & \cr 
(\nu/\nu^{SC}_{bm})^{1/2} & \nu^{SC}_{bc}<\nu<\nu^{SC}_{bm} \cr & \cr 
(\nu/\nu^{SC}_{bm})^{(2-s)/2} & \nu>\nu^{SC}_{bm}}\right. \ .
\end{equation}
\begin{equation}\label{Fnu_syc_SC3}
{\nu F^{SC}_{\nu}(t_{\rm sd}>t_{3})\over Y_b\,\nu_{bc} F_{\nu_{bc}}}=\left\{\matrix{
(\nu^{SC}_{bm}/\nu^{SC}_{bc})^{(3-s)/2}(\nu^{SC}_{bsa}/\nu^{SC}_{bm})^{4/3}(\nu/\nu^{SC}_{bsa})^2 & 
\ \ \nu<\nu^{SC}_{bsa} \cr & \cr 
(\nu^{SC}_{bm}/\nu^{SC}_{bc})^{(3-s)/2}(\nu/\nu^{SC}_{bm})^{4/3} & 
\nu^{SC}_{bsa}<\nu<\nu^{SC}_{bm} \cr & \cr 
(\nu/\nu^{SC}_{bc})^{(3-s)/2} & \nu^{SC}_{bm}<\nu<\nu^{SC}_{bc} \cr & \cr 
(\nu/\nu^{SC}_{bc})^{(2-s)/2} & \nu>\nu^{SC}_{bc}}\right. \ .
\end{equation}
where $\nu^{SC*}_{bsa}\equiv\max(1,\gamma_{bc}^2)\nu_{bsa}$, and 
\begin{eqnarray}\label{SCcase1}
\nu^{SC}_{bc}& = & \gamma_{bc}^2\nu_{bc}\approx  
2.1\times 10^{6}\,(1+z)^{-1}a^{-2}\eta_{2/3}^{-7/2}\epsilon_{be,1/3}^{2}\epsilon_{bB,-3}^{-3/2}
 E_{\rm rot,53}^{-7/2}\beta_{b,-1}^{21/2}t_{\rm sd,0}^{13/2}\;{\rm Hz}\ ,
\nonumber \\
\nu^{SC}_{bm}& = & \gamma_{bm}^2\nu_{bm}\approx  
6.4\times 10^{20}\,(1+z)^{-1}\eta_{2/3}^{9/2} \epsilon_{be,1/3}^4\epsilon_{bB,-3}^{1/2}
\xi_{e,1/3}^{-4}
E_{\rm rot,53}^{1/2}\gamma_{w,4.5}^4\beta_{b,-1}^{-3/2}t_{\rm sd,0}^{-3/2}\;{\rm Hz}\ ,
\end{eqnarray}
and
\begin{equation}\label{SC_SSA}
\nu^{SC}_{bsa}=\left\{\matrix{
\gamma_{bc}^2\nu_{bsa}\approx
{1.4\times 10^{10}\over(1+z)}a^{-7/6}\eta_{2/3}^{-2}\epsilon_{be,1/3}^{-7/6} 
\epsilon_{bB,-3}^{-5/6}\xi_{e,1/3}^{1/3}
E_{\rm rot,53}^{-5/3}\gamma_{w,4.5}^{-1/3}\beta_{b,-1}^{16/3}
t_{\rm sd,0}^{3}\;{\rm Hz} & \ \ t_1<t_{\rm sd}<t_2 \cr & \cr
\gamma_{bc}^2\nu_{bsa}\approx {8.0\times 10^{12}\over(1+z)}a^{-1/2}
\eta_{2/3}^{-4/5}\epsilon_{be,1/3}^{-1/5}
\epsilon_{bB,-3}^{-3/10}\xi_{e,1/3}^{3/5} E_{\rm rot,53}^{-1/5}\gamma_{w,4.5}^{-3/5}
\beta_{b,-1}^{6/5}t_{\rm sd,0}^{1/5}\;{\rm Hz} & \ \ t_2<t_{\rm sd}<t_3 \cr & \cr
\gamma_{bm}^2\nu_{bsa}\approx {3.6\times 10^{16}\over(1+z)}\eta_{2/3}^{6/5}
\epsilon_{be,1/3}\epsilon_{bB,-3}^{1/5}
\xi_{e,1/3}^{-2/5}E_{\rm rot,53}^{4/5}\gamma_{w,4.5}^{2/5}
\beta_{b,-1}^{-9/5}t_{\rm sd,0}^{-9/5}\;{\rm Hz} & \ \ t_{\rm sd}>t_3 }\right. \ .
\end{equation}
The peak of $\nu F^{SC}_\nu$ is simply $Y_b$ times the peak of the synchrotron $\nu F_\nu$,
which for fast cooling is given by
\begin{equation}\label{nu_F_numax_FC_syn}
\nu_{bm} F_{\nu_{bm}}=\nu_{bm}F_{\nu,{\rm max}}\sqrt{\nu_{bc}\over\nu_{bm}}\approx
5.1\times 10^{-15}a^{-1/2}\eta_{2/3}\epsilon_{be,1/3}^{1/2}\epsilon_{bB,-3}^{1/2}
E_{\rm rot,53}t_{\rm sd,0}^{-1}d_{L28}^{-2}\ ,
\end{equation}
and for slow cooling is given by this expression multiplied by the factor $a$.

\subsection{The Klein-Nishina Effect}
\label{plerion_KN}

When the energy of the seed (synchrotron) photon in the rest frame of the 
scattering electron exceeds the electrons rest energy, 
$\gamma_e h\nu^{syn}\gtrsim m_e c^2$, then we move from the Thomson limit to 
the Klein-Nishina (KN) regime, where there is a reduction in the scattering cross section 
and the electron recoil becomes important. The corresponding frequencies where the 
KN limit is reached for the seed synchrotron photon and upscattered (SC) photon
are given by
\begin{equation}\label{KN_freq}
{\nu_{KN}^{SC}\over\gamma_e}=\gamma_e\nu_{KN}^{syn}
=\nu_{KN,0}\equiv{m_e c^2\over h}={c\over\lambda_c}=1.23\times 10^{20}\;{\rm Hz}\ ,
\end{equation}
where $\lambda_c$ is the Compton wavelength. As the energy increase of the photon 
due to the scattering cannot exceed that of the electron, the photon energy of 
$h\nu_{KN}^{SC}=\gamma_e m_e c^2$ sets the natural upper limit to the frequency 
of photons that are upscattered by electrons with a LF $\gamma_e$. Typically, either the
fractional energy gain in successive scatterings is $Y\sim\tau_T\gamma_e^2\ll 1$,
or the KN limit is reached for the second scattering, $\gamma_e^3h\nu^{syn}\gtrsim m_e c^2$,
thus allowing us to ignore multiple scatterings.

In KN effects typically become important only at $\nu^{SC}>{\rm max}
(\nu^{SC}_{bsa},\nu_{bc}^{SC})$ in the fast cooling regime, or at $\nu>\nu_{bm}^{SC}$
in the slow cooling regime. We therefor restrict our discussion to these frequency ranges
of the SSC spectrum.
Within this frequency range, the SSC 
flux density at a given frequency consists of roughly equal contributions from seed 
synchrotron photons that extend over a finite range of frequencies, that are upscattered by 
electrons within a finite range of LFs $\gamma_b$, so that $\gamma_b^2\nu^{syn}=\nu^{SC}
={\rm const}$. For this reason, a significant change in the SSC flux at a certain frequency 
$\nu^{SC}$, due to KN effects, will occur only when all the electrons that contribute
significantly to this frequency reach the KN limit. Since $\nu_{KN}^{SC}(\gamma_b)\propto
\gamma_b$, this occurs when the electron with the maximal $\gamma_b$ that still contributes 
significantly to $\nu^{SC}$, $\gamma_{max}(\nu^{SC})$, reaches the KN limit. 
For fast cooling
\begin{equation}\label{gamma_max(nu)}
\gamma_{max}(\nu)=\left\{\matrix{\sqrt{\nu/\nu_{bp}}\ \ &
\gamma_{bc}^2\nu_{bp}<\nu<\gamma_{bm}^2\nu_{bp} \cr & \cr 
\gamma_{bm}\ \ & \gamma_{bm}^2\nu_{bp}<\nu<\nu_{bm}^{SC} \cr & \cr
\sqrt{\nu/\nu_{bm}}\ \ & \nu_{bm}^{SC}<\nu<\gamma_{bM}^2\nu_{bm} \cr & \cr
\gamma_{bM}\ \ & \nu>\gamma_{bM}^2\nu_{bm}}\right. \ ,
\end{equation}
where $\nu_{bp}\equiv{\rm max}(\nu_{bsa},\nu_{bc})$, while for slow cooling
\begin{equation}\label{gamma_max(nu)2}
\gamma_{max}(\nu)=\left\{\matrix{\sqrt{\nu/\nu_{bm}}\ \ &
\nu_{bm}^{SC}<\nu<\gamma_{bc}^2\nu_{bm} \cr & \cr 
\gamma_{bc}\ \ & \gamma_{bc}^2\nu_{bm}<\nu<\nu_{bc}^{SC} \cr & \cr
\sqrt{\nu/\nu_{bc}}\ \ & \nu_{bc}^{SC}<\nu<\gamma_{bM}^2\nu_{bc} \cr & \cr
\gamma_{bM}\ \ & \nu>\gamma_{bM}^2\nu_{bc}}\right. \ ,
\end{equation}
where $\nu_{bsa}<\nu_{bm}$. 
The KN effects become important at the frequency $\nu^{SC}_{kN,1}$ that 
satisfies\footnote{from equations (\ref{gamma_max(nu)}) and (\ref{gamma_max(nu)2}) 
it is evident that there is always exactly one solution to this equation, while the 
explicit form of this solution depends on the frequency range in which it is obtained.}
$\nu=\gamma_{max}(\nu)\nu_{KN,0}$. For fast cooling
\begin{equation}\label{nu_KN1_FC}
\nu^{SC}_{kN,1}=\left\{\matrix{\nu_{KN,0}^2/\nu_{bp}\ \ &
\gamma_{bc}^2\nu_{bp}<\nu^{SC}_{kN,1}<\gamma_{bm}^2\nu_{bp} \cr & \cr 
\gamma_{bm}\nu_{KN,0}\ \ & \gamma_{bm}^2\nu_{bp}<\nu^{SC}_{kN,1}<\nu_{bm}^{SC} \cr & \cr
\nu_{KN,0}^2/\nu_{bm}\ \ & \nu_{bm}^{SC}<\nu^{SC}_{kN,1}<\gamma_{bM}^2\nu_{bm} \cr & \cr
\gamma_{bM}\nu_{KN,0}\ \ & \nu^{SC}_{kN,1}>\gamma_{bM}^2\nu_{bm}}\right. \ .
\end{equation}
while for slow cooling
\begin{equation}\label{nu_KN1_SC}
\nu^{SC}_{kN,1}=\left\{\matrix{\nu_{KN,0}^2/\nu_{bm}\ \ &
\nu_{bm}^{SC}<\nu^{SC}_{kN,1}<\gamma_{bc}^2\nu_{bm} \cr & \cr 
\gamma_{bc}\nu_{KN,0}\ \ & \gamma_{bc}^2\nu_{bm}<\nu^{SC}_{kN,1}<\nu_{bc}^{SC} \cr & \cr
\nu_{KN,0}^2/\nu_{bc}\ \ & \nu_{bc}^{SC}<\nu^{SC}_{kN,1}<\gamma_{bM}^2\nu_{bc}\ \cr & \cr
\gamma_{bM}\nu_{KN,0}\ \ & \nu^{SC}_{kN,1}>\gamma_{bM}^2\nu_{bc}}\right. \ .
\end{equation}

At $\nu<\nu^{SC}_{kN,1}$ KN effects are unimportant and the SSC spectrum is given by 
equations (\ref{Fnu_syc_SC1}-\ref{Fnu_syc_SC3}). At $\nu>\nu^{SC}_{kN,1}$ KN effects 
become important. In order to calculate $F_{\nu}$ in this range we 
first need to estimate the scattering optical depth $\tau_e(\gamma_b)$ of electrons with 
LF $\gamma_e\sim\gamma_b$, which is of the same order as the total optical depth of 
electrons with $\gamma_e>\gamma_b$:
\begin{equation}\label{tau_e}
\tau_e(\gamma_b)\approx\sigma_T R_b\int_{\gamma_b}^{\gamma_{bM}}d\gamma_e N(\gamma_e)\ ,
\end{equation}
where $N(\gamma_e)$ is given in equations (\ref{N(gam)_FC}) and (\ref{N(gam)_SC}) for 
fast cooling and for slow cooling, respectively. For fast cooling we obtain
\begin{equation}\label{tau_e_FC}
\tau_e(\gamma_e)\approx\tau_T\times\left\{\matrix{
\gamma_{bc}/\gamma_e\ \ & \gamma_{bc}<\gamma_e<\gamma_{bm} \cr & \cr
\gamma_{bc}\gamma_{bm}^{s-1}\gamma_e^{-s}\ \ & \gamma_{bm}<\gamma_e<\gamma_{bM}}\right.\ ,
\end{equation}
while for slow cooling 
\begin{equation}\label{tau_e_SC}
\tau_e(\gamma_e)\approx\tau_T\times\left\{\matrix{
(\gamma_e/\gamma_{bm})^{1-s}\ \ & \gamma_{bm}<\gamma_e<\gamma_{bc} \cr & \cr
\gamma_{bc}\gamma_{bm}^{s-1}\gamma_e^{-s}\ \ & \gamma_{bc}<\gamma_e<\gamma_{bM}}\right.\ .
\end{equation}
At $\nu>\nu^{SC}_{kN,1}$ the SSC flux density is dominated by the contribution from 
electrons with $\nu_{KN}^{SC}(\gamma_b)\sim\nu$ and is given by 
\begin{equation}\label{F_nu_KN}
F_{\nu}^{SC}\approx\tau_e(\gamma_e=\nu/\nu_{KN,0})F^{syn}_{\nu_{KN,0}^2/\nu}\ .
\end{equation}
The frequency dependence arising from the first term on the r.h.s of equation 
(\ref{F_nu_KN}) is just the $\gamma_e$ dependence of $\tau_e(\gamma_e)$ [that appears 
explicitly in equations (\ref{tau_e_FC}) and (\ref{tau_e_SC})], while the 
second term introduces a frequency dependence of $\nu^{-\beta}$ where $F_{\nu}^{syn}
\propto \nu^{\beta}$ at $\nu^{syn}=\nu_{KN,0}^2/\nu^{SC}$. 

In all cases the largest frequency that the SSC spectrum reaches 
is ${\rm min}[\nu_{KN}^{SC}(\gamma_{bM}),\gamma_{bM}^2\nu_{bM}]$.
For $\nu_{KN}^{SC}(\gamma_{bM})>\gamma_{bM}^2\nu_{bM}$ the SSC spectrum is given by 
equations (\ref{Fnu_syc_SC1}-\ref{Fnu_syc_SC3}), with no changes. For $\gamma_{bM}^2
{\rm max}(\nu_{bm},\nu_{bc})<\nu_{KN}^{SC}(\gamma_{bM})<\gamma_{bM}^2\nu_{bM}$,
the SSC spectrum is the same as in equations (\ref{Fnu_syc_SC1}-\ref{Fnu_syc_SC3})
up to $\nu_{KN}^{SC}(\gamma_{bM})$ where it sharply ends [note that 
$\nu_{KN}^{SC}(\gamma_{bM})=\nu_{KN,1}^{SC}$ in this region]. 

For fast cooling with $\gamma_{bm}^2\nu_{bp}<\nu_{KN,1}^{SC}<\gamma_{bM}^2\nu_{bm}$ we have 
$F^{SC}_{\nu}\propto\nu^{-\beta-s}$ for $\nu_{KN,1}^{SC}<\nu<\nu_{KN}^{SC}(\gamma_{bM})$, 
and the spectrum ends at $\nu_{KN}^{SC}(\gamma_{bM})$. Immediately above $\nu_{KN,1}^{SC}$ 
we have $\beta=-1/2$ while $\beta$ might change to $1/3$ (for $\nu_{bc}>\nu_{bsa}$) or $2$ 
(for $\nu_{bc}<\nu_{bsa}$) at a higher frequency, $\nu_{KN,0}^2/\nu_{bp}$, producing a 
spectral break at this frequency, if $\nu_{KN,0}^2/\nu_{bp}<\nu_{KN}^{SC}(\gamma_{bM})$ 
[i.e. if $\gamma_{bM}^2\nu_{bp}>\nu_{KN}^{SC}(\gamma_{bM})$]. For fast cooling with 
$\gamma_{bc}^2\nu_{bp}<\nu_{KN,1}^{SC}<\gamma_{bm}^2\nu_{bp}$ we have $F^{SC}_{\nu}
\propto\nu^{-\beta-1}$ for $\nu_{KN,1}^{SC}<\nu<\nu_{KN}^{SC}(\gamma_{bm})$, and 
$F^{SC}_{\nu}\propto\nu^{-\beta-s}$ for $\nu_{KN}^{SC}(\gamma_{bm})<\nu<\nu_{KN}^{SC}
(\gamma_{bM})$, where for $\nu_{bsa}>\nu_{bc}$ we have $\beta=2$, while for 
$\nu_{bc}>\nu_{bsa}$ we have $\beta=1/3$ immediately above $\nu_{KN,1}^{SC}$, which may 
change to $\beta=2$ at $\nu_{KN,0}^2/\nu_{bsa}$ if $\nu_{KN,0}^2/\nu_{bsa}<\nu_{KN}^{SC}
(\gamma_{bM})$.

For slow cooling with $\gamma_{bc}^2\nu_{bm}<\nu_{KN,1}^{SC}<\gamma_{bM}^2\nu_{bc}$ we have 
$F^{SC}_{\nu}\propto\nu^{-\beta-s}$ for $\nu_{KN,1}^{SC}<\nu<\nu_{KN}^{SC}(\gamma_{bM})$, 
and the spectrum ends at $\nu_{KN}^{SC}(\gamma_{bM})$. Immediately above $\nu_{KN,1}^{SC}$ 
we have $\beta=(1-s)/2$ while $\beta$ might change to $1/3$ at a higher frequency, $\nu_{KN,0}^2/\nu_{bm}$, producing a spectral break at this frequency, if $\nu_{KN,0}^2/\nu_{bm}<
\nu_{KN}^{SC}(\gamma_{bM})$. For slow cooling with $\gamma_{bc}^{SC}<\nu_{KN,1}^{SC}<
\gamma_{bc}^2\nu_{bm}$ we have $F^{SC}_{\nu}\propto\nu^{1-s-\beta}$ for 
$\nu_{KN,1}^{SC}<\nu<\nu_{KN}^{SC}(\gamma_{bc})$, and $F^{SC}_{\nu}\propto\nu^{-s-\beta}$ 
for $\nu_{KN}^{SC}(\gamma_{bc})<\nu<\nu_{KN}^{SC}(\gamma_{bM})$, where $\beta=1/3$ 
immediately above $\nu_{KN,1}^{SC}$, and may change to $\beta=2$ at 
$\nu_{KN,0}^2/\nu_{bsa}$ if $\nu_{KN,0}^2/\nu_{bsa}<\nu_{KN}^{SC}(\gamma_{bM})$.

\subsection{Opacity to Pair Production}
\label{pair_opacity}

High energy photons emitted in the PWB, either by the plerion itself or by the prompt
GRB or afterglow that occur inside the PWB, may interact with lower energy photons 
of the strong radiation field of the plerion to create $e^\pm$ pairs. For sufficiently
high photon energies, the optical depth to this process, $\tau_{\gamma\gamma}$, may 
exceed unity, so that they could not escape and reach the observer. We now calculate 
the photon energy $\epsilon$ (in units of $m_ec^2$) for which 
$\tau_{\gamma\gamma}(\epsilon)=1$. This sets the maximal photon energy that will not 
be effected by this process. 

The radiation field of the plerion is roughly homogeneous and isotropic 
within the largest radius where the radiation is emitted, which we parameterize
as $fR_b$. For a fast cooling PWB ($t_{\rm sd}\ll t_3\sim 65\;{\rm yr}$), the radiation 
is emitted within a thin layer behind the wind termination shock, at the radius $R_s$, 
so that $f=R_s/R_b$  For an adiabatic bubble, as the one considered here,
the value of this ratio ranges between 0.2 and 0.5 (KG). For a slow cooling PWB ($t_{\rm sd}>t_3\sim 65\;{\rm yr}$) the
radiation is emitted from the whole volume of the PWB, and $f=1$. The internal shocks 
that produce the prompt GRB emission take place at a radius smaller than $fR_b$, and the 
energy density of the plerion radiation field is 
$U_{ph}\approx a\xi_{e}E_{\rm rot}/t_{\rm sd}2\pi(fR_b)^2 c$.
However, the relevant target photons for pair production 
with high energy photons are the synchrotron photons, 
since the synchrotron component is dominant at low energies.
Therefore, we should use the energy density of the synchrotron photons, 
$U_{syn}=U_{ph}/(1+Y_b)$.
As we shall see in \S \ref{AG_EC}, the afterglow emission typically occurs at $R>fR_b$,
where the radiation field is not homogeneous, but rather drops as $U_{ph}\propto R^{-2}$, 
and is not isotropic, causing a smaller typical angle between the trajectories of the two photons
that could possibly produce an $e^\pm$ pair. Both effects reduce $\tau_{\gamma\gamma}$ for the 
afterglow emission, compared to that for the prompt GRB or the plerion emission itself, that we 
calculate below.

The number density of synchrotron photons, $n_\epsilon$, per unit 
dimensionless photon energy $\epsilon$, may be obtained from the shape of the 
synchrotron spectrum, and its normalization is
\begin{equation}\label{U_ph}
{a \xi_{e} E_{\rm rot}\over c\, t_{\rm sd}2\pi (fR_b)^2(1+Y_b)}=
U_{syn}=m_ec^2\int d\epsilon\, n_\epsilon\epsilon=gm_ec^2
\epsilon_{bm}^2n_{\epsilon_{bm}}\ ,
\end{equation}
where $U_{ph}$ is the photon energy density of the plerion and $g=2(s-1)/(s-2)$.
The optical depth to pair production is given by
\begin{equation}\label{tau_gg}
\tau_{\gamma\gamma}(\epsilon)\approx \sigma_T fR_b \epsilon^{-1}n_{1/\epsilon}=
\sigma_T fR_b {U_{syn}\over gm_ec^2}\epsilon_{bm}^{(s-2)/2}\epsilon^{s/2}\ ,
\end{equation}
and $\tau_{\gamma\gamma}(\epsilon)=1$ is satisfied for
\begin{equation}\label{nu_gg}
h\nu_{\gamma\gamma}={\epsilon\, m_e c^2\over(1+z)} \approx 
{2.4\;{\rm GeV}\over(1+z)}\,\left({f_{1/3}\over \sqrt{a}}\right)^{10\over 11}
\eta_{2/3}^{-5/22}\epsilon_{be,1/3}^{3/11}
\epsilon_{bB,-3}^{-1/2}\xi_{e,1/3}^{-8/11}E_{\rm rot,53}^{-21/22}
\gamma_{w,4.5}^{-2/11}\beta_{b,-1}^{23/22}t_{\rm sd,0}^{43/22} \ ,
\end{equation}
where $f_{1/3}=f/(1/3)$.
It can be seen from equation (\ref{nu_gg}), that opacity to pair production becomes 
important only at very high photon energies, and is larger for smaller $t_{\rm sd}$.

For an elongated PWB, the radiation field can generally have a rather different 
structure, resulting in a different expression for $\nu_{\gamma\gamma}$. However,
one can imagine a simple scenario where the structure of the radiation field is
similar to the spherical case, and equation (\ref{nu_gg}) is still applicable with
simple substitutions. This can occur if the PWB is fast cooling and most of the 
radiation is emitted just behind the wind termination shock, and the latter is 
roughly spherical, with a radius, $R_s$, similar to the equatorial radius, $R_{\rm eq}$. 
In this case we should make the usual substitution $\beta_b=3R_{\rm eff}/2ct_{\rm sd}$, 
as well as $f=R_s/R_{\rm eff}\sim R_{\rm eq}/R_{\rm eff}$. For example, with 
$\beta_b\approx 1$ and $R_{\rm eq}/R_{\rm eff}\sim 0.1$ this would increase 
$\nu_{\gamma\gamma}$ by a factor of $\sim 3$.

\subsection{Prospects for Direct Detection}
\label{plerion_dd}

An important prediction of this model is a strong radiation field within 
the PWB. We now examine the possibility of directly observing the radiation 
emanating from the PWB during the time between the SN and the GRB events. 
For time separations $t_{\rm sd}$ smaller than
\begin{equation}\label{t_tau}
t_\tau=0.4\,\left({M_{\rm SNR}\over 10M_\odot}\right)^{1/2}
\beta_{b,-1}^{-1}\;{\rm yr}\ ,
\end{equation}
the SNR shell has a Thomson optical depth larger than unity, and would 
therefore obscure the radiation emitted within the PWB. For $t_{\rm sd}<t_\tau$
the emission due to the radioactive decay of Ni and Co in the SNR shell,
might be observed, as in a regular supernova. However, even at the peak
of the supernova emission, it will be hard to detect at a cosmological distance. 
This difficulty is also present in ongoing searches for high redshift supernovae, 
where like in our case, random patches of the sky need to be searched, as there 
is no prompt GRB and afterglow emission to tell us where and when to look.

If there is considerable clumping of the SNR shell before this time, then equation 
(\ref{t_tau}) gives the time when the average optical depth of the SNR 
equals 1, while for regions of the shell with a less than average density 
the optical depth can drop below unity at a somewhat earlier time.
This constraint may also be eased if the geometry of the PWB is not spherical
(an elongated PWB), and the mean density of the SNR shell is significantly smaller 
near the poles compared to near the equator, and our line of sight is near one of 
the poles (as is required in order to see the prompt $\gamma$-ray emission from a 
jetted GRB).

Once the SNR shell becomes optically thin, the PWB radiation may be 
detected if the flux that arrives at the observer is sufficiently large. 
For simplicity, we calculate the flux at the time of the GRB explosion,
$t_{\rm sd}$ after the supernova event, since the relevant quantities scale as 
power laws with the time $t$ after the supernova, and the system spends most of 
its (logarithmic) time near $t=t_{\rm sd}$.
For concreteness, we consider the observed flux at the radio, optical and 
X-ray, for a typical frequency in each of these frequency ranges: 
$\nu_{\rm rad}=5\times 10^{9}\,{\rm Hz}$, $\nu_{\rm op}=5\times 10^{14}\,
{\rm Hz}$ and $\nu_{\rm X}=10^{18}\,{\rm Hz}$. 

For the radio we have $\nu_{\rm rad}<\nu_{bsa}$ for $t_{\rm sd}<t_{sa,{\rm rad}}$, 
$\nu_{bsa}<\nu_{\rm rad}<\nu_{bc}$ for $t_{sa,{\rm rad}}<t_{\rm sd}<t_{c,{\rm rad}}$ and
$\nu_{bc}<\nu_{\rm rad}<\nu_{bm}$ for $t_{\rm sd}>t_{c,{\rm rad}}$, where
\begin{eqnarray}
t_{sa,{\rm rad}} &=& 3.9\,(1+z)^{-1}\epsilon_{be,1/3}^{-1/6}\epsilon_{bB,-3}^{1/6}
\xi_{e,1/3}^{1/3}E_{\rm rot,53}^{1/3}\gamma_{w,4.5}^{-1/3}\beta_{b,-1}^{-2/3}
\nu_{9.7}^{-1}\;{\rm yr}\ ,
\label{t_sa,rad}\\
t_{c,{\rm rad}} &=& 20\,(1+z)^{2/5}\eta_{2/3}^{3/5}\epsilon_{be,1/3}^{2/5}\epsilon_{bB,-3}^{1/5}
E_{\rm rot,53}^{3/5}\beta_{b,-1}^{-9/5}\nu_{14.7}^{2/5}\;{\rm yr}\ ,
\label{t_c,rad}
\end{eqnarray}
and
\begin{equation}\label{F_op2}
F_{\nu}^{syn}=\left\{\matrix{
{34\;\mu{\rm Jy}\over(1+z)^{-3}}\,\eta_{2/3}^{-1/4}\epsilon_{be,1/3}^{-1/12}\epsilon_{bB,-3}^{-1/6}
\xi_{e,1/3}^{1/6}E_{\rm rot,53}^{-1/12}\gamma_{w,4.5}^{-1/6}
\beta_{b,-1}^{29/12}t_{\rm sd,0}^{9/4}\nu_{9.7}^{2}d_{L28}^{-2}
& \ \ t_{\rm sd}<t_{sa,rad} \cr & \cr 
{1.0\;{\rm mJy}\over(1+z)^{-1/2}}\,\eta_{2/3}^{-1/4}\epsilon_{be,1/3}^{-1/2}\epsilon_{bB,-3}^{1/4}
\xi_{e,1/3}E_{\rm rot,53}^{3/4}\gamma_{w,4.5}^{-1}\beta_{b,-1}^{3/4}t_{\rm sd,0}^{-1/4}
\nu_{9.7}^{-1/2}d_{L28}^{-2} & \ \ t_{sa,rad}<t_{\rm sd}<t_{c,rad} \cr & \cr
{0.50\;{\rm Jy}\over(1+z)^{-4/3}}\,\eta_{2/3}\epsilon_{be,1/3}^{1/3}\epsilon_{bB,-3}^{2/3}
\xi_{e,1/3}E_{\rm rot,53}^2\gamma_{w,4.5}^{-1}\beta_{b,-1}^{-3}t_{\rm sd,0}^{-7/3}\nu_{9.7}^{1/3}
d_{L28}^{-2} & \ \ t_{\rm sd}>t_{c,rad}}\right. \ .
\end{equation}

For $t_{\rm sd}<t_{m,{\rm op}}$ we have $\nu_{\rm op}<\nu_{bm}$, while for 
$t_{\rm sd}>t_{m,{\rm op}}$, the ordering is reversed $\nu_{op}>\nu_{bm}$, where 
$t_{m,op}$ is given by 
\begin{equation}\label{t_bm,op}
t_{m,{\rm op}}=0.22\,(1+z)^{-2/3}\eta_{2/3}^{5/3}\epsilon_{be,1/3}^{4/3}\epsilon_{bB,-3}^{1/3}
\xi_{e,1/3}^{-4/3}E_{\rm rot,53}^{1/3}\gamma_{w,4.5}^{4/3}
\beta_{b,-1}^{-1}\nu_{14.7}^{-2/3}\;{\rm yr}\ .
\end{equation}
The optical flux is dominated by synchrotron emission and is given by
\begin{equation}\label{F_op}
F_{\nu}^{syn}=\left\{\matrix{
{3.1\;\mu{\rm Jy}\over(1+z)^{-1/2}}\,\eta_{2/3}^{-1/4}\epsilon_{be,1/3}^{-1/2}\epsilon_{bB,-3}^{1/4}
\xi_{e,1/3}E_{\rm rot,53}^{3/4}\gamma_{w,4.5}^{-1}
\beta_{b,-1}^{3/4}t_{\rm sd,0}^{-1/4}\nu_{14.7}^{-1/2}d_{L28}^{-2}
& \ \ t_{\rm sd}<t_{m,{\rm op}} \cr & \cr 
{0.82\;\mu{\rm Jy}\over(1+z)^{1/10}}\,\eta_{2/3}^{5/4}\epsilon_{be,1/3}^{7/10}\epsilon_{bB,-3}^{11/20}
\xi_{e,1/3}^{-1/5}E_{\rm rot,53}^{21/20}\gamma_{w,4.5}^{1/5}
\beta_{b,-1}^{-3/20}t_{\rm sd,0}^{-23/20}\nu_{14.7}^{-11/10}d_{L28}^{-2}
& \ \ t_{\rm sd}>t_{m,{\rm op}}}\right. \ .
\end{equation}

The X-ray is typically above $\nu_{bm}$, and for the synchrotron emission we have
\begin{equation}\label{F_X_syn}
\nu F_{\nu}^{syn}={1.9\times 10^{-15}\over(1+z)^{1/10}}\,\eta_{2/3}^{5/4}
\epsilon_{be,1/3}^{7/10}\epsilon_{bB,-3}^{11/20}\xi_{e,1/3}^{-1/5}
E_{\rm rot,53}^{21/20}\gamma_{w,4.5}^{1/5}
\beta_{b,-1}^{-3/20}t_{\rm sd,0}^{-23/20}\nu_{18}^{-1/10}
d_{L28}^{-2}\;{{\rm erg}\over{\rm cm^2\,s}}\ .
\end{equation}
For sufficiently small values $t_{\rm sd}$ we have $\nu_{bc}^{SC}<\nu_{\rm X}<\nu_{bm}^{SC}$. 
As $t_{\rm sd}$ increases, then $\nu_{bm}^{SC}$ decreases below the X-ray for 
$t_{\rm sd}>t_{m,{\rm X}}^{SC}$), 
and $\nu_{bc}^{SC}$ grows above the X-ray for $t_{\rm sd}>t_{c,{\rm X}}^{SC}$,
where the relative ordering of $t_{m,{\rm X}}^{SC}$ and $t_{c,{\rm X}}^{SC}$
depends on the values of the other parameters. For our fiducial values we have 
\begin{eqnarray}
t_{m,{\rm X}}^{SC} &=& 25\,(1+z)^{-2/3}\eta_{2/3}^{3}\epsilon_{be,1/3}^{8/3}\nu_{bm}^{SC}\epsilon_{bB,-3}^{1/3}
\xi_{e,1/3}^{-8/3}E_{\rm rot,53}^{1/3}\gamma_{w,4.5}^{8/3}
\beta_{b,-1}^{-1}\nu_{18}^{-2/3}\;{\rm yr}\ ,
\label{t_c,SC}\\
t_{c,{\rm X}}^{SC} &=& 80\,(1+z)^{2/13}a^{4/13}\eta_{2/3}^{7/13}\epsilon_{be,1/3}^{4/13}
\epsilon_{bB,-3}^{3/13}E_{\rm rot,53}^{7/13}\beta_{b,-1}^{-21/13}\nu_{18}^{2/13}\;{\rm yr}\ .
\label{t_m,SC}
\end{eqnarray}
For this ordering of these two times we have $\nu_{\rm X}>\max(\nu_{bm}^{SC},\nu_{bc}^{SC})$ for 
$t_{m,{\rm X}}^{SC}<t_{\rm sd}<t_{c,{\rm X}}^{SC}$, and $\nu_{bm}^{SC}<\nu_{\rm X}<\nu_{bc}^{SC}$ 
for $t_{\rm sd}>t_{c,{\rm X}}^{SC}$. For the other ordering, $t_{c,{\rm X}}^{SC}<t_{m,{\rm X}}^{SC}$,
we have $\nu_{\rm X}<\min(\nu_{bm}^{SC},\nu_{bc}^{SC})$ for 
$t_{c,{\rm X}}^{SC}<t_{\rm sd}<t_{m,{\rm X}}^{SC}$ and $\nu_{bm}^{SC}<\nu_{\rm X}<\nu_{bc}^{SC}$
for $t_{\rm sd}>t_{m,{\rm X}}^{SC}$. The X-ray is always below the KN limit, and the SSC $\nu F_{\nu}$, 
is given by
\begin{equation}\label{F_X_SSC}
{\nu F_{\nu}^{SC}\over \left({\rm erg\over cm^{2}\,s}\right)}=
\end{equation}
$$
\left\{\matrix{
{3.7\cdot 10^{-15}\over(1+z)^{-1/2}}\eta_{2/3}^{-5/4}\epsilon_{be,1/3}^{-1}
\epsilon_{bB,-3}^{-1/4}\xi_{e,1/3}^{2}E_{\rm rot,53}^{3/4}\gamma_{w,4.5}^{-2}
\beta_{b,-1}^{7/4}t_{\rm sd,0}^{-1/4}\nu_{18}^{1/2}d_{L28}^{-2}
& \ \ t_{\rm sd}<\min(t_{c,{\rm X}}^{SC},t_{m,{\rm X}}^{SC}) \cr & \cr 
{1.8\cdot 10^{-13}\over(1+z)^{1/10}}\eta_{2/3}^{29/20}\epsilon_{be,1/3}^{7/5}
\epsilon_{bB,-3}^{1/20}\xi_{e,1/3}^{-2/5}E_{\rm rot,53}^{21/20}\gamma_{w,4.5}^{2/5}
\beta_{b,-1}^{17/20}t_{\rm sd,0}^{-23/20}\nu_{18}^{-1/10}d_{L28}^{-2} 
& \ \ t_{m,{\rm X}}^{SC}<t_{\rm sd}<t_3 \cr & \cr
{2.8\cdot 10^{-14}\over(1+z)^{1/10}}\eta_{2/3}\epsilon_{be,1/3}^{16/15}
\epsilon_{bB,-3}^{-11/180}\xi_{e,1/3}^{-8/45}E_{\rm rot,53}^{5/6}\gamma_{w,4.5}^{8/45}
\beta_{b,-1}^{31/60}t_{\rm sd,0}^{-127/180}\nu_{18}^{-1/10}d_{L28}^{-2} 
& \ \ t_3<t_{\rm sd}<t_{c,{\rm X}}^{SC} \cr & \cr
{2.0\cdot 10^{-5}\over(1+z)^{-4/3}}\,\eta_{2/3}^{5/3}\epsilon_{be,1/3}^{2/3}
\epsilon_{bB,-3}\xi_{e,1/3}^2E_{\rm rot,53}^{11/3}\gamma_{w,4.5}^{-2}
\beta_{b,-1}^{-7}t_{\rm sd,0}^{-17/3}\nu_{18}^{4/3}d_{L28}^{-2}
& \ \ t_{c,{\rm X}}^{SC}<t_{\rm sd}<t_3 \cr & \cr
{3.0\cdot 10^{-10}\over(1+z)^{-4/3}}\,\eta_{2/3}^{-1}\epsilon_{be,1/3}^{-4/3}
\epsilon_{bB,-3}^{1/3}\xi_{e,1/3}^{10/3}E_{\rm rot,53}^{7/3}\gamma_{w,4.5}^{-10/3}
\beta_{b,-1}^{-4}t_{\rm sd,0}^{-3}\nu_{18}^{4/3}d_{L28}^{-2}
& \ \ t_3<t_{\rm sd}<t_{m,{\rm X}}^{SC} \cr & \cr
{1.2\cdot 10^{-7}\over(1+z)^{-2/5}}\,\eta_{2/3}^{16/5}\epsilon_{be,1/3}^{12/5}
\epsilon_{bB,-3}^{4/5}\xi_{e,1/3}^{-2/5}E_{\rm rot,53}^{14/5}\gamma_{w,4.5}^{2/5}
\beta_{b,-1}^{-22/5}t_{\rm sd,0}^{-22/5}\nu_{18}^{2/5}d_{L28}^{-2} 
& \ \ t_{\rm sd}>\max(t_{c,{\rm X}}^{SC},t_{m,{\rm X}}^{SC})}\right. \ .
$$

The emission from the PWB calculated above is at the time of the GRB explosion, 
$t_{\rm sd}$. However, an important question one needs to address is for how long 
after the onset of the GRB explosion will the radiation from the plerion persist.
Once the SMNS collapses to a black hole, giving rise to the GRB event, the pulsar 
wind stops abruptly, and no new electrons, freshly accelerated at the wind 
termination shock, are injected into the plerion from this point on. The electrons
in the PWB begin to cool radiatively, and adiabatic cooling of the electrons 
become important on the dynamical time of the plerion, $t_{\rm sd}$ (where this is
a good estimate of the dynamical time also for an elongated PWB). Once the last 
accelerated electrons cool below the Lorentz factor at which they radiate at some 
observed frequency $\nu$, no more radiation is emitted at that frequency. This cooling 
time is given by
\begin{equation}\label{t_cool}
t_{\rm cool}=8.9\,(1+z)^{-1/2}\eta_{2/3}^{-3/4}
\epsilon_{be,1/3}^{-1/2}\epsilon_{bB,-3}^{-1/4}\beta_{b,-1}^{9/4}
E_{\rm rot,53}^{-3/4}t_{\rm sd,0}^{9/4}\nu_{9.7}^{-1/2}\;{\rm days}\ ,
\end{equation}
where the above numerical coefficient is for the radio, while for the optical and 
X-ray the numerical coefficient is 40 min and $24\;{\rm s}$, respectively.
For our simple example of an elongated PWB, where $\beta_b\approx 1$, the cooling
time is about two orders of magnitude larger.
Due to the strong dependence on $t_{\rm sd}$, $t_{\rm cool}$ can become quite 
large for large values of $t_{\rm sd}$, especially for radio frequencies. 

A possibly more severe constraint arises from the geometrical time delay in the 
arrival of photons to the observer, from the different parts of the PWB
\begin{equation}\label{t_geom}
t_{\rm g}\sim {R_b\over c}=
24.2\,\beta_{b,-1}t_{\rm sd,0}\;{\rm days}\ .
\end{equation}
This is to say, that even if the emission from the PWB would stop at once, with the
onset of the GRB, the radiation would still reach the observer for $t_{\rm g}$ after 
the GRB, just due to the geometrical time delay in the arrival of photons to the 
observer from the far side of the PWB, compared to the side facing the observer.
For $t_{\rm sd}<t_3$, where $t_3$ is typically rather large [see equation 
(\ref{t_3})], the PWB is in the fast cooling regime, and most of the emission occurs 
near the radius of the termination shock, $R_s$. This reduces $t_{\rm g}$ by a factor
of $R_s/R_b$. However, this is not expected to account for more than a factor of 
$\sim 5$ (KG). For an elongated PWB we have $t_{\rm g}\sim R_s/c$ for $t_{\rm sd}\ll t_3$,
while for $t_{\rm sd}>t_3$ we have $t_{\rm g}\sim R_p/c\approx t_{\rm sd}$.
The emission from the PWB should persist for an observed time of
$t_{\rm pl}\sim(1+z)\max[t_{\rm g},\min(t_{\rm sd},t_{\rm cool})]$
after the GRB, which as can be seen from equations (\ref{t_cool}) and (\ref{t_geom})
should be at least a few days after the GRB, but can also be much larger 
($t_{\rm pl}\gtrsim t_{\rm g}$, see equation \ref{t_geom}). 

\begin{figure}
\centering
\noindent
\includegraphics[width=12cm]{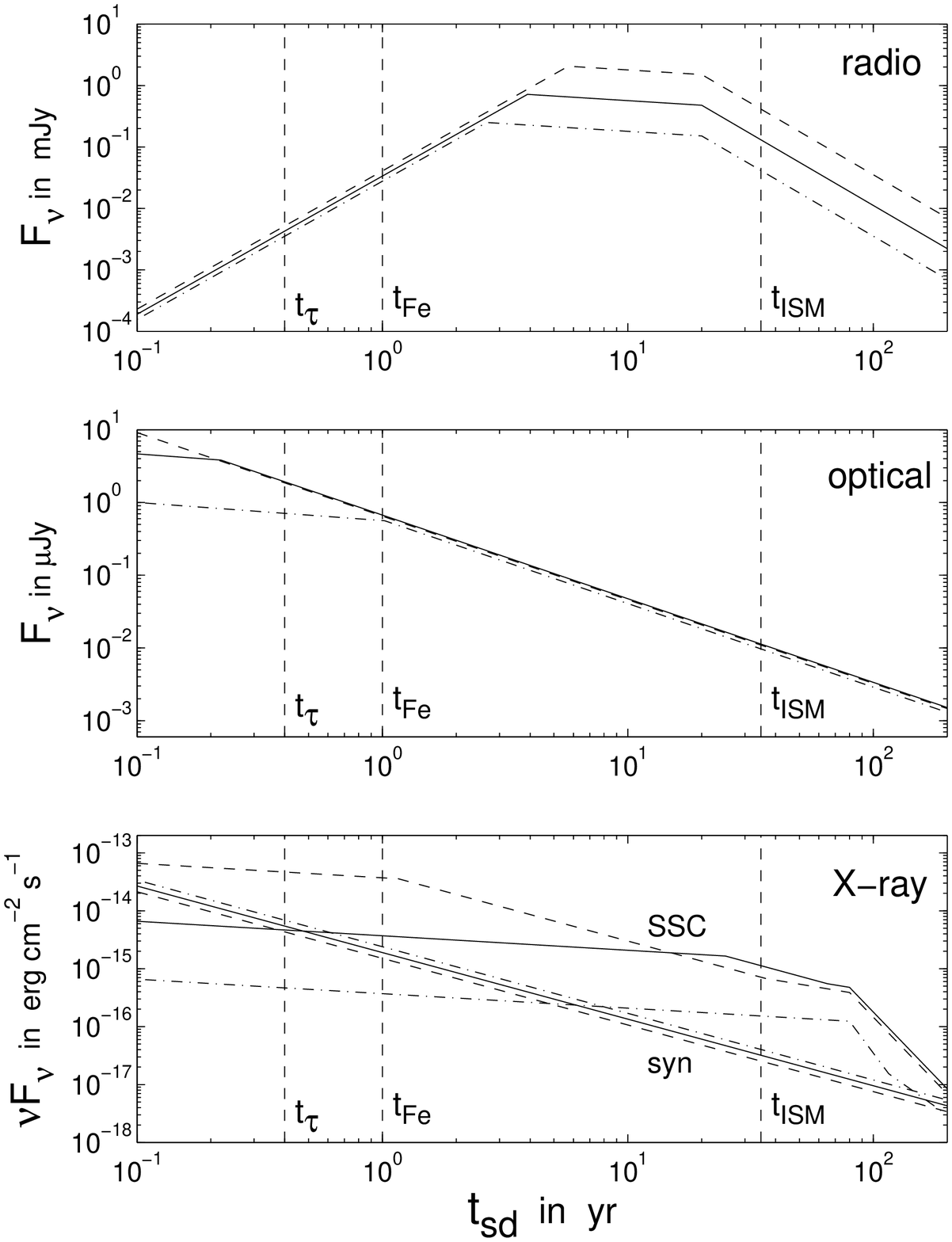}
\caption{\label{fig1} 
The flux density, $F_\nu$, 
of the plerion emission at the time of the GRB (i.e. a time $t_{\rm sd}$ 
after the supernova event), as a function of $t_{\rm sd}$.
The three panels show the flux density in the radio, optical and X-ray bands
($\nu=5\times 10^9,\,5\times 10^{14},\,10^{18}\;{\rm Hz}$, respectively).
For the radio and optical we show $F_\nu$ of the synchrotron emission,
and for the X-ray we show $\nu F_\nu$ for both the synchrotron and SSC components.
The dashed, solid and dot-dashed lines represent $\log_{10}(\gamma_w)=4,\,4.5$ and $5$,
respectively. Dashed lines are shown at $t_\tau$ (below which the Thomson optical depth 
is larger than unity), $t_{\rm Fe}$ (below which iron line features can appear in the X-ray 
spectrum of the afterglow) and $t_{\rm ISM}$ (for which the effective density of the PWB is 
similar to that of a typical ISM, i.e. $1\;{\rm cm^{-3}}$.
}
\end{figure}

The plerion emission at the radio, optical and X-ray bands is shown in Figure \ref{fig1}.
For reference, we also show the times $t_\tau$ (below which the Thomson optical depth 
is larger than unity), $t_{\rm Fe}$ (below which iron line features can appear in the X-ray 
spectrum of the afterglow) and $t_{\rm ISM}$ (for which the effective density of the PWB is 
similar to that of a typical ISM, i.e. $1\;{\rm cm^{-3}}$). In the radio, the typical limiting 
flux for detection is $\sim 0.1\;{\rm mJy}$, and upper limits at this flux level, at a time
$t<t_{\rm pl}$ after the GRB, would exclude $2\lesssim t_{\rm sd,0}\lesssim 20$ for 
$\gamma_w\lesssim 10^5$, while for $2\lesssim t_{\rm sd,0}\lesssim 65$ this would imply
$\gamma_w\gtrsim 10^4$. Values of $t_{\rm sd}\lesssim 2\;{\rm yr}$ or $t_{\rm sd}\gtrsim 65\;{\rm yr}$
are hard to constrain with the radio. 

Optical upper limits at the R-band at the level of $\sim 24-25$th magnitude 
($F_\nu\lesssim 0.5{\rm \mu Jy}$) would imply $\gamma_w\gtrsim 10^5$ for
$t_{\rm sd}<t_{\rm Fe}$, i.e. for afterglows with iron lines.
More stringent upper limits, may provide more severe constraints. For example, an upper
limit of $0.01\;{\rm \mu Jy}$ (i.e. R=28.6 which might be reached with HST)
implies $t_{\rm sd}\gtrsim t_{\rm ISM}$.

In the X-ray, for $\gamma_w\lesssim 10^{4.5}$ the SSC emission dominates over the 
synchrotron emission for all $t_{\rm sd}\gtrsim t_\tau$, and for $\gamma_w=10^5$
it dominates for $t_{\rm sd}\gtrsim 7\;{\rm yr}$. The typical limiting flux for detection 
in the X-ray is a few $\times 10^{-14}\;{\rm erg\;cm^{-2}\;s^{-1}}$, and upper limits at this
level may imply $\gamma_w\gtrsim 10^4$ for $t_{\rm sd}\lesssim t_{\rm Fe}$. For 
$t_{\rm sd}\gtrsim t_{\rm Fe}$, such upper limits cannot provide any useful constraints.

\section{Effects on the Prompt GRB Emission}
\label{prompt_GRB}

The prompt gamma-ray emission is believed to arise from internal shocks 
within the GRB outflow, due to variability in its Lorentz factor $\Gamma$
(Rees \& M\'esz\'aros 1994; Sari \& Piran 1997). In order for this process to 
be efficient, it needs to occur before the ejecta is significantly decelerated 
by the ambient medium. Therefore, the main effect that a plerionic environment 
may have on the prompt GRB stage is through inverse Compton scattering of the 
photons from the external plerion radiation field (which we shall refer to as 
external Compton, or EC). 

The external (plerion) radiation field in the local 
rest frame of the shocked shells is $U'_{ph,ext}=\Gamma^2 U_{ph,ext}$ 
where $U_{ph,ext}$ is given by equation (\ref{U_ph}). The electrons 
radiatively cool by the combination of the synchrotron, SSC and EC 
processes the timescales for which are (in the comoving frame)
$t'_{\rm syn}\sim 6\pi m_e c/\sigma_T B'^2 \gamma$, $t'_{SC}=t'_{\rm syn}/Y$ 
and $t'_{EC}=t'_{\rm syn}/X$, where $Y$ is the Compton y-parameter and 
\begin{equation}\label{X_IS}
X={U'_{ph,ext}\over \epsilon_B e'}=3.1\times 10^{-4}a^{1/2}f_{1/3}^{-2}
\xi_{e,1/3}\epsilon_{be,1/3}^{-1/2}\epsilon_{bB,-3}^{1/2}
\epsilon_{e}\epsilon_{B}^{-1}\beta_{b,-1}^{-2}t_{\rm sd,0}^{-3}
E_{\rm rot,53}L_{52}^{-1}\Gamma_{2.5}^{8}t_{v,-2}^{2}\ ,
\end{equation}
is the ratio of the energy density of the external radiation field
and the magnetic field in the local rest frame of the shocked shells, which is 
also the ratio between the energies in the EC and synchrotron components. 
As can be seen from equation (\ref{X_IS}), $X\ll 1$ for typical parameters. 
In order to have $X\gtrsim 1$, we need $t_{\rm sd}\lesssim 1\;{\rm yr}$ and
$\Gamma\gtrsim 10^3$. For an elongated PWB of the type described just before 
\S \ref{plerion_syn}, with $f\sim R_s/R_{\rm eff}\sim 0.1$ and 
$\beta_b\approx 1$, $X$ is roughly the same.

The EC component is due to the scattering of external photons from the plerion 
radiation field by the electrons from the GRB ejecta. This scattering can be 
done either by hot (relativistic) electrons, or by cold (non-relativistic) 
electrons, the latter being either in cold shells or cold portions of colliding 
shells (in either regions before the shock, or at a distance larger than 
$\gamma_c\Delta'$ behind the shock for $\gamma_c<1$). 

For the colliding shells, we assume that Thomson optical depth, $\tau_T$, is smaller than 1. 
We provide detailed expressions for one representative plerion spectrum, the one given in equation 
(\ref{Fnu_syc_VFC}), that is relevant for $t_{\rm sd}<t_2\sim 12\:{\rm yr}$ (see equation \ref{t_2}), 
which is of most interest. Similar expressions for the other plerion spectra can be readily derived 
in a similar manner. The plerion SSC emission in this regime has a peak for the $\nu F_\nu$ at 
$\nu_{bm}^{SC}\gtrsim 10^{20}\;{\rm Hz}$ for $t_{\rm sd}\lesssim 1\;{\rm yr}$, and will 
therefore be above the KN cutoff for both hot and cold electrons in the outflowing shells, 
and its contribution to the EC emission can be neglected. The resulting EC spectrum due
to scattering by the hot electrons is 
\begin{equation}\label{Fnu_EC_IS}
{\nu F_{\nu}^{EC}\over \nu_m F_{\nu_m}}=X\times\left\{\matrix{
(\nu_{sa}^{EC}/\nu^{EC}_{m})^{1/2}(\nu/\nu_{sa}^{EC})^2 & 
\nu<\nu_{sa}^{EC} \cr & \cr 
(\nu/\nu^{EC}_{m})^{1/2} & \nu_{sa}^{EC}<\nu<\nu^{EC}_{m}\cr & \cr
(\nu/\nu^{EC}_{m})^{1-s/2} & \nu^{EC}_{m}<\nu<\nu_{KN}^{EC}(\gamma_{M})}
\right. \ .
\end{equation}
where $\nu_{sa}^{EC}\equiv\Gamma^2\max(\gamma_c^2,1)\nu_{bsa}$ and
\begin{eqnarray}\label{freq_IS_EC}
\nu_{m}^{EC}&=&\Gamma^2\gamma_m^2\nu_{bm}\approx 1.9\times 10^{23}\eta_{2/3}^{5/2}
\epsilon_{be,1/3}^{2}\epsilon_{bB,-3}^{1/2}\xi_{e,1/3}^{-2}E_{\rm rot,53}^{1/2}
\gamma_{w,4.5}^{2}\beta_{b,-1}^{-3/2}t_{\rm sd,0}^{-3/2}\epsilon_e^{2}\Gamma_{2.5}^{2}\;{\rm Hz}\ ,
\nonumber \\
\nu_{KN,1}^{SC}&=&\nu_{KN,0}^2/\nu_{bm}\approx 3.0\times 10^{27}\eta_{2/3}^{-5/2}
\epsilon_{be,1/3}^{-2}\epsilon_{bB,-3}^{-1/2}\xi_{e,1/3}^{2}E_{\rm rot,53}^{-1/2}
\gamma_{w,4.5}^{-2}\beta_{b,-1}^{-3/2}t_{\rm sd,0}^{3/2}\;{\rm Hz}\ ,
\nonumber \\
\nu_{KN}^{SC}(\gamma_M)&=&\Gamma\gamma_{M}\nu_{KN,0}\approx 6.6\times 10^{27}
(1+Y+X)^{-1/2}\epsilon_e^{1/4}\epsilon_B^{-1/4}L_{52}^{-1/4}
\Gamma_{2.5}^{5/2}t_{v,-2}^{1/2}\;{\rm Hz}\ .
\end{eqnarray}
If $\nu_{KN,1}^{SC}<\nu_{KN}^{SC}(\gamma_{M})$ then we have $\nu F_\nu\propto\nu^{1/2-s}$
for $\nu_{KN,1}^{SC}<\nu<\nu_{KN}^{SC}(\gamma_{M})$. The peak of the $\nu F_\nu^{EC}$ spectrum, 
of the EC component from hot electrons is, is $\sim 10^{-10}(X/10^{-4})\;
{\rm erg\,cm^{-2}\,s^{-1}}$ and is a factor of $X\sim 10^{-4}$ 
(for $t_{\rm sd}\sim 1\;{\rm yr}$) smaller than that of the synchrotron 
component, and therefore might be detected only for extreme 
parameters ($t_{\rm sd,0}\lesssim1$, $\Gamma\gtrsim 10^3$).

For the scattering by cold 
electrons, the optical depth is approximately  
\begin{equation}\label{tau_T_IS}
\tau_T=\sigma_t n'_e\Delta'\approx 0.02\epsilon_e^{-1}L_{52}
\Gamma_{2.5}^{-5}t_{v,-2}^{-1}\ ,
\end{equation}
and the Compton y-parameter is 
$Y=\min(\tau_T,1)\Gamma^2$. This implies that all frequencies of the plerion 
spectrum are shifted upwards by a factor of $\Gamma^2$, and the corresponding 
flux density, $F_{\nu}$, should be multiplied by $\min(\tau_T,1)$. 
The only exception to this simple re-scaling is that the spectral slope below 
the upscattered self absorption frequency will be $F_\nu\propto\nu$, instead of 
the $\nu^2$ in the plerion spectrum. The peak of the $\nu F_\nu^{EC}$ spectrum will be at
\begin{equation}\label{nu_mEC_IS}
\Gamma^2\nu_{bm}\approx 5.1\times 10^{18}\,(1+z)^{-1}\eta_{2/3}^{5/2}\epsilon_{be,1/3}^2
\epsilon_{bB,-3}^{1/2}\xi_{e,1/3}^{-2} E_{\rm rot,53}^{1/2}\gamma_{w,4.5}^2\beta_{b,-1}^{-3/2}
t_{\rm sd,0}^{-3/2}\;{\rm Hz}\ ,
\end{equation}
which is typically at the hard X-ray or soft gamma-ray for $t_{\rm sd}\lesssim 1\;{\rm yr}$.
However, the peak of $\nu F_\nu^{EC}$ for this component is a factor of $Y\leq\Gamma^2\sim 10^5$
larger than $\nu_{bm}F_{\nu_{bm}}\lesssim 10^{-14}\;{\rm erg\,cm^{-2}\,s^{-1}}$ and is therefore
$\lesssim 10^{-9}\;{\rm erg\,cm^{-2}\,s^{-1}}$, and will typically hide below the standard GRB 
emission.

Another possible effect of the plerion radiation field is that photons with energy $\gtrsim 1 \, t_{\rm sd,0}^2 \;{\rm GeV}$
cannot escape the emission region due to a large opacity to pair 
production ($\tau_{\gamma\gamma}>1$). Therefore, all the components of the prompt GRB emission,
including synchrotron, SSC and EC, will have an upper cutoff at this photon energy.

Finally, we consider the effect of the Compton drag due to the plerion radiation field
on the GRB outflow\footnote{The emission from the initial supernova itself always 
contributes much less to the total radiation field inside the PWB, and may therefore 
be neglected.}. The effect of Compton drag in GRBs was considered in the context of
the collapsar model, where the radiation comes from the walls of a funnel along the
rotational axis of the progenitor star (Ghisellini et al. 2000; Lazzati et al. 2000). 
The rate of energy loss of a shell of initial Lorentz factor $\Gamma_0$
rest mass $M$ and solid angle $\Omega_j$, is given by
\begin{equation}\label{dEdt_compdrag}
{dE\over dt}={d\Gamma\over dt}Mc^2=-\Omega_j R^2 c\Gamma^2 U_{ph}\min(\tau_T,1)
\end{equation}
where $t\approx R/c$ is the lab frame time, $\tau_T=(R/R_\tau)^{-2}$ is the 
Thomson optical depth of the shell, $R_\tau$ is the radius where this optical 
depth drops below unity\footnote{We have used the total photon energy 
density of the PWB, $U_{ph}$, that includes the SSC component, even though for 
$t_{\rm sd}\lesssim 30\;$yr, most of the energy in the SSC component is in 
photons that are above the Klein-Nishina limit, and would therefore have 
a reduced cross section for scattering. Since we show that Compton drag 
is unimportant even without taking into account the reduced 
cross section, this effect can only strengthen our conclusion.}. 
We render Eq. \ref{dEdt_compdrag}
dimensionless by introducing $\tilde{R}\equiv R/R_\tau=\tau_T^{-1/2}$,
\begin{equation}\label{dinvgdR_compdrag}
{d(1/\Gamma)\over d\tilde{R}}=A\min(\tilde{R}^2,1)\quad,\quad
A\equiv{\Omega U_{ph} R_\tau^3\over Mc^2}
\end{equation}
This gives
\begin{equation}\label{Gamma_compdrag}
{\Gamma(\tilde{R})\over\Gamma_0}=\left\{\matrix{ (1+A\Gamma_0\tilde{R}^3/3)^{-1}  & 
\tilde{R}<1\ \   \cr & \cr
\left[1+A\Gamma_0(\tilde{R}-2/3)\right]^{-1}  & \tilde{R}>1\ \ }\right. \ ,
\end{equation}
where
\begin{eqnarray}\label{R_tau}
R_\tau &=& 3.3\times 10^{12}\,\epsilon_\gamma^{-1/2}L_{52}^{1/2}\Gamma_{2.5}^{-1/2}t_{v,-2}^{1/2}\;{\rm cm}\ ,
\nonumber \\ \label{A}
A &=& 1.6\times 10^{-8}a\, f_{1/3}^{-2}\epsilon_\gamma^{-1/2}\xi_{e,1/3}E_{\rm rot,53}
\beta_{b,-1}^{-2}t_{\rm sd,0}^{-3}L_{52}^{1/2}\Gamma_{2.5}^{-1/2}t_{v,-2}^{1/2}\ ,
\end{eqnarray}
where $\epsilon_\gamma$ is the fraction of the kinetic luminosity of the GRB outflow that is 
converted into the gamma-ray emission. As can be seen from Eqs. (\ref{Gamma_compdrag}) and
(\ref{A}), $A\Gamma_0\ll 1$ and therefore $\Gamma(R_\tau)\cong\Gamma_0$, while for 
$R\gg R_\tau$ the fractional change in $\Gamma$ is given by $A\Gamma_0\tilde{R}$.
If the radius, $R_{\rm cd}$, at which deceleration due to Compton drag becomes significant,
is larger than the deceleration radius, $R_{\rm dec}\approx f R_b$, due to the sweeping 
up ofthe PWB material, then the deceleration due to Compton drag is at most comparable
(and never dominant) to the deceleration due to the ambient medium, for\footnote{
This is since the Lorentz factor decreases with radius as $R^{-1}$ due to Compton drag 
and as $R^{(k-3)/2}$ due to the ambient medium.} $k\leq 1$. Therefore, Compton drag will 
have a significant effect on the decelleration only if $R_{\rm cd}<R_{\rm dec}$, 
which for our fiducial values may be expressed as $t_{\rm sd}<0.3 f^{1/2}\;$yr.
For such low values of $t_{\rm sd}$ the SNR shell is still optically thick
to Compton scattering ($t_{\rm sd}<t_\tau$), so that we do not expect to
see the GRB or afterglow emission. We conclude that the deceleration of
the GRB ejecta due to Compton drag is negligible for relevant values of $t_{\rm sd}$.

\section{Effects on the Afterglow Emission}
\label{Afterglow}

At a time $t_{\rm sd}$ after the supernova event, the SMNS collapses and triggers 
the GRB explosion, sending a fireball and relativistic blast wave into the
PWB. When the GRB ejecta has swept up enough of the outlying material, it is decelerated,
and it drives a strong relativistic shock into the external medium, that is responsible 
for the afterglow (AG) emission.
 
In this section we study the observational consequences of the plerionic environment
inside the PWB, that are different from the standard ``cold'', weakly magnetized 
proton-electron external medium. One of the advantages of having the PWB as the 
environment for the GRB afterglow is that it naturally yields high values of 
$\epsilon_e$ and $\epsilon_B$ (the fraction of the internal energy in the electrons 
and in the magnetic field, respectively) behind the AG shock (KG). High values of 
$\epsilon_e$ are expected from the fact that relativistic pulsar-type winds are likely 
dominated by an electron-positron component, whereas significant values of $\epsilon_B$ 
should naturally occur if the winds are characterized by a high magnetization parameter. 
We expect $\epsilon_B\sim\epsilon_{bB}$, and use the same fiducial value for these two 
parameters. The electrons in the PWB are typically colder than the protons by the time 
the afterglow shock arrives, and most of the energy is in the internal energy of the 
hot protons. This might suggest that $\epsilon_e$ can be slightly smaller than
$\epsilon_{be}$ and motivates us to use $\epsilon_e=\epsilon_{e,-1}/10$ for our 
fiducial values.

The values of the physical quantities behind the AG shock can be determined from the 
appropriate generalizations of the hydrodynamic conditions used in the case of a ``cold'' 
medium taking into account the fact that the preshock gas is now ``hot'' and should be well 
described by a relativistic equation of state, $p=e/3=w/4$, where $w$ is the enthalpy density 
and $p$ is the particle pressure. In the following we largely follow the analysis of KG .
The deceleration of the AG shock is determined by the total enthalpy of the external medium, 
$w_{tot}$, which includes contributions from the particles and the magnetic field enthalpy, 
$B^2/4\pi$, where the latter contribution is negligible for our choice of parameters 
($\epsilon_B\ll 1$). This make it convenient to define an ``equivalent'' hydrogen number 
density $n_{\rm H, equiv} \equiv w_{\rm tot}/m_p c^2\approx w/m_p c^2=(4/3)e/m_p c^2$, 
in analogy with the traditional parameterization of the external medium enthalpy density, 
$w= n_{\rm H} m_p c^2$, that is relevant for a standard ISM or stellar-wind environment. 

In general both the energy and the electron number density 
may be function of the distance $r$ from the center of the PWB
and can be parametrized as
\begin{eqnarray}
e(r) &=& A_e r^{-k_*}\label{e_r} 
\quad ; \quad  A_e =\frac{(3-k_*)\eta E_{\rm rot}}{4\pi R_{b}^{3-k}}\ ,\\\label{n_r}
n_e(r) &=& A_n r^{-k} \quad ; \quad A_n =\frac{(3-k)N_e}{4\pi R_{b}^{3-k}}\ ,
\end{eqnarray}
where $N_e=\dot{N}_e t_{\rm sd}$ is the total number of the electrons in the PWB
and $\dot{N}_e$ is given in Eq. \ref{Ndot}.
When a large fraction of the energy density in the PWB goes to the proton component we have
$\eta\sim 1$ and we can expect both $n_{\rm H, equiv}$ and $n_e$ to have a similar radial 
dependence, i.e. $k=k_*$. The expected values of $k$ typically ranges between $k=0$, similar 
the the ISM, and $k=1$, that is intermediate between an ISM and a stellar wind (KG).
For an elongated PWB, things can get much more complicated, since $e$ and $n_e$ can 
depend not only on $r$ but also on the angle $\theta$ from the polar axis. However, if
the $\theta$ dependence within the opening angle of the GRB jet is small, and the 
dependence on $r$ may be reasonably approximated by a power law, then our formalism 
still holds for $k\approx 0$, with the usual substitution of 
$\beta_b=3R_{\rm eff}/2ct_{\rm sd}$ (see the beginning of \S \ref{PlerEmis}).
For $k>0$ we also need to change the normalization in equation 
(\ref{e_r}) and (\ref{n_r}) accordingly.

The expressions for the radius $R_{AG}$ and the Lorentz factor $\Gamma_{AG}$ 
of the shocked AG material can be derived using energy conservation
\begin{equation}
E_{\rm iso}=\Gamma^2_{AG}\int^{R_{AG}}_{R_S} n_{\rm H, equiv}
m_p c^2 4\pi r^2 dr\, ,
\end{equation}
where $E_{\rm iso}$ is the isotropic equivalent energy of the AG shock, and using the relation
\begin{equation}
t\sim \frac{R_{AG}}{4 c\Gamma_{AG}^2}\ ,
\end{equation}
where $t$ is the observed time. We obtain
\begin{equation}\label{Rag}
R_{AG} = 
\left[\frac{3 E_{\rm iso} c t R_b^{3-k}}
{\eta\,E_{\rm rot}} \right]^{1/(4-k)} 
\end{equation}
\begin{equation}\label{gammaAC}
\Gamma_{AG} = \sqrt{\frac{R_{AG}}{4ct}}
\propto t^{(k-3)\over 2(4-k)}\propto \left\{\matrix{ t^{-3/8} & \quad k=0  \cr & \cr 
t^{-1/3} & \quad k=1 }\right. \ .
\end{equation}
The postshock energy and particle density (in the shock comoving frame) are given by
\begin{equation}
e^\prime  =  4 \Gamma^2_{AG}w  \quad , \quad
n_e^\prime  =  4\Gamma_{AG} n_e
\end{equation}
The electron distribution is assumed to be similar to that of internal shocks,
$N(\gamma)\propto \gamma^{-p}$ for $\gamma_m<\gamma<\gamma_M$, and we use $p=2.5$ 
to obtain the numerical values. 

As mentioned in \S \ref{pair_opacity}, the plerion radiation field is roughly homogeneous 
and isotropic within the the radius $fR_b$ where the plerion emission takes place.
At $r\lesssim fR_b$ the external (plerion) photon energy density, $U_{ph,ext}$,
is given by equation (\ref{U_ph}), and we may use the relation 
$U'_{ph,ext}=\Gamma_{AG}^2U_{ph,ext}$ (that is valid for an isotropic radiation field) 
to obtain
\begin{equation}\label{X_AG}
X={U'_{ph,ext}\over \epsilon_B e'}\approx\left\{\matrix{
2.1\,a^{1/2}\,f_{1/3}^{-2}\eta_{2/3}^{-1}\epsilon_{be,1/3}^{-1/2}
\epsilon_{bB,-3}^{1/2}\epsilon_{B,-3}^{-1}\xi_{e,1/3}\beta_{b,-1} 
& k=0  \cr & \cr 
1.8\,a^{1/2}\,f_{1/3}^{-2}\eta_{2/3}^{-4/3}\epsilon_{be,1/3}^{-1/2}
\epsilon_{bB,-3}^{1/2}\epsilon_{B,-3}^{-1}\xi_{e,1/3}
E_{\rm rot,53}^{-1/3}\beta_{b,-1}^{2/3}t_{sd,1.5}^{-1/3}
E_{\rm iso,53}^{1/3}t_{\rm days}^{1/3} & k=1
}
\right. \ ,
\end{equation}
where $t_{sd,1.5}=t_{sd}/10^{1.5}$. For $t_{\rm sd}>t_3\sim 65\;{\rm yr}$ the PWB is slow 
cooling, and $f=1$, so that $R_{AG}<fR_b$ throughout the afterglow.
For $t_{\rm sd}<t_3$ the plerion emission is radiated 
within a shell of width $R_b-R_s$ times the ratio, $(t_{\rm sd}/t_3)^2$, of the cooling time 
of electrons with $\gamma_{bm}$, and the dynamical time $t_{\rm sd}$. This implies that
generally, $f=\min\{1,(R_s/R_b)[1-(t_{\rm sd}/t_3)^2]+(t_{\rm sd}/t_3)^2\}$.
For $t_{\rm sd}<t_3$ we have $f<1$ and therefore $R_{AG}<fR_b$ only at sufficiently early times
after the GRB. For $t_{\rm sd}\ll t_3$, the radiation is emitted within a
thin shell behind the wind termination shock, at $R_s$, and $f\approx R_s/R_b$.
In this case $R_{AG}>fR_b$ throughout the afterglow. We study the implications
in the following.

At $r>fR_b$, the plerion radiation field is no longer isotropic or homogeneous,
and we model the plerion radiation field  as resulting from emission by a 
uniformly bright sphere with a radius $fR_b$, and obtain
\begin{equation}\label{U_ph_AG}
U_{ph}={\xi_e L_w \over 2\pi c f R_b^2}(1-\mu)\approx {\xi_e L_w\over 4\pi c\, r^2}\quad ,\quad
U'_{ph}={\xi_e L_w \over 2\pi c f R_b^2}{\Gamma_{AG}^2\over 3\beta}[(1-\beta\mu)^3-(1-\beta)^3]\ ,
\end{equation}
where $\mu\equiv[1-(fR_b/r)^2]^{1/2}$ and $\beta=(1-\Gamma_{AG}^{-2})^{1/2}$. The ratio of the 
photon energy in the local and the observer frames is now
\begin{equation}\label{U_ph_ratio}
{U'_{ph}\over U_{ph}}={\Gamma_{AG}^2\left[(1-\beta\mu)^3-(1-\beta)^3\right]\over 
3\beta(1-\mu)}\approx\left\{\matrix{
\Gamma_{AG}^2(1-\mu)^2/3\approx(fR_b/r)^4\Gamma_{AG}^2/12 & r\ll\Gamma_{AG}fR_b \cr & \cr 
(1-\beta)/(1+\beta)\approx 1/4\Gamma_{AG}^2 & r\gg\Gamma_{AG}fR_b 
}\right. \ .
\end{equation}
During the early afterglow, $R_{AG}$ is relatively small and $\Gamma_{AG}\gg 1$, so that 
$R_{AG}\ll\Gamma_{AG}fR_b$, and the first limit of equation (\ref{U_ph_ratio}) is applicable,
implying
\begin{equation}\label{X_1}
X_1\approx\left\{\matrix{
0.26\,a^{-1/2}f_{1/3}^4\eta_{2/3}^{1/2}\epsilon_{be,1/3}^{-1/2}
\epsilon_{bB,-3}^{1/2}\epsilon_{B,-3}^{-1}\xi_{e,1/3}
E_{\rm rot,53}^{3/2}\beta_{b,-1}^{5/2}t_{sd,1.5}^{3/2}
E_{\rm iso,53}^{-3/2}t_{\rm days}^{-3/2} & k=0 \cr & \cr
0.93\,a^{-1/2}f_{1/3}^4\eta_{2/3}^{2/3}\epsilon_{be,1/3}^{-1/2}
\epsilon_{bB,-3}^{1/2}\epsilon_{B,-3}^{-1}\xi_{e,1/3}
E_{\rm rot,53}^{5/3}\beta_{b,-1}^{8/3}t_{sd,1.5}^{5/3}
E_{\rm iso,53}^{-5/3}t_{\rm days}^{-5/3} & k=1
}\right. \ .
\end{equation}
During the course of the afterglow its radius increases while its Lorentz factor decreases,
so that eventually $R_{AG}$ becomes larger than $\Gamma_{AG}fR_b$, and the second limit of
equation (\ref{U_ph_ratio}) becomes relevant, implying
\begin{equation}\label{X_2}
X_2\approx\left\{\matrix{
1.3\times 10^{-4}\,a^{-1/2}\epsilon_{be,1/3}^{-1/2}
\epsilon_{bB,-3}^{1/2}\epsilon_{B,-3}^{-1}\xi_{e,1/3}E_{\rm rot,53}
t_{sd,1.5}^{3/2}E_{\rm iso,53}^{-1}t_{\rm days} & k=0 \cr & \cr
2.0\times 10^{-4}\,a^{-1/2}\epsilon_{be,1/3}^{-1/2}\epsilon_{bB,-3}^{1/2}
\epsilon_{B,-3}^{-1}\xi_{e,1/3}E_{\rm rot,53}
t_{sd,1.5}^{3/2}E_{\rm iso,53}^{-1}t_{\rm days} & k=1
}\right. \ ,
\end{equation}
as long as the afterglow shock is still relativistic.
One can combine the two limits and use $X=\max(X_1,X_2)$. 
However, since the region where these asymptotic expressions for $X$
are not a very good approximation may play an important role, we use equation 
(\ref{U_ph_AG}) rather than equations (\ref{X_1}) and (\ref{X_2}) for all our calculations.

\begin{figure}
\centering
\noindent
\includegraphics[width=14cm]{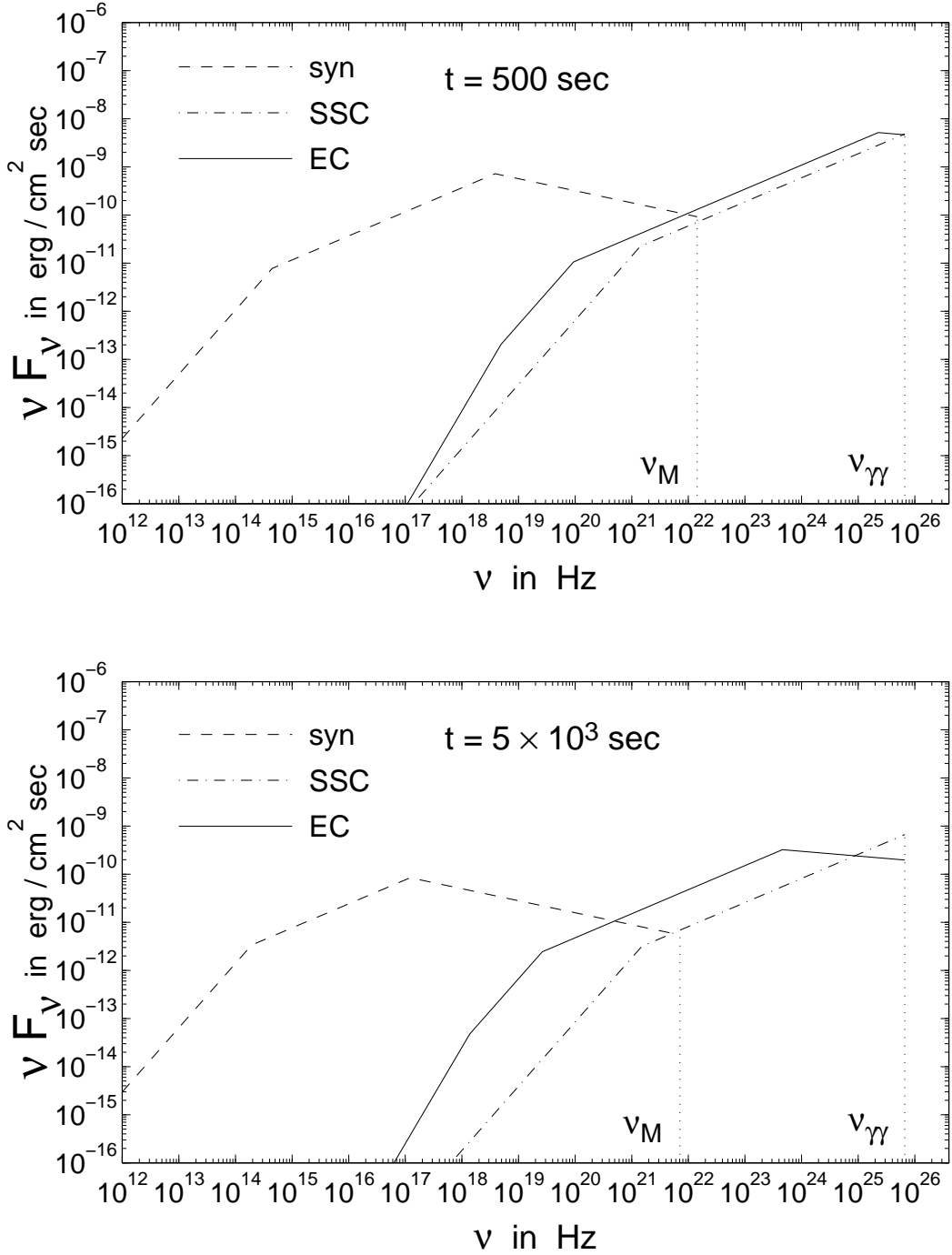}
\caption{\label{fig2} 
The afterglow $\nu F_\nu$ spectrum at $t=500\;{\rm s}$ (upper panel) and 
$5\times 10^3\;{\rm s}$ (lower panel) after the GRB, for our fiducial parameters, 
and for $t_{\rm sd}=20\;{\rm yr}$, $E_{\rm rot,53}=0.5$, $z=1$, $R_s/R_b=0.1$. 
Dotted vertical lines indicate $\nu_M$ where the upper cutoff for the synchrotron 
emission is located, and $\nu_{\gamma\gamma}$ where the upper cutoff of the SSC 
and EC (due to pair opacity) is located.
}
\end{figure}

It is also worth to note that the average shift in frequency of the photons between the observer 
frame and local and rest frame is $\langle \nu'/\nu\rangle=\Gamma_{AG}[1-\beta(1+\mu)/2]
\approx[1+\Gamma_{AG}^2(1-\mu)^2]/2\Gamma_{AG}$, and varies between $\Gamma_{AG}/2$
and $1/2\Gamma_{AG}$. This should be compared to the usual factor of $\Gamma_{AG}$
for an isotropic (plerion) radiation field, and implies lower typical EC frequencies,
by a factor of $[1-\beta(1+\mu)/2]$. For simplicity we do not include this factor 
in the expressions for the EC frequencies, but we do take it into account in Figure \ref{fig2}, 
and when deriving constraints on the model parameters.

The electron cooling time is $t_{\rm syn}/(1+Y+X)$ where
the Compton y-parameter may be obtained by solving the equation 
\begin{equation}\label{Y_AG1}
Y\approx\tau_T a \gamma_c\gamma_m\approx {a\epsilon_e\over\epsilon_B(1+Y+X)}\ ,
\end{equation}
which gives (Granot \& K\"onigl 2001):
\begin{equation}\label{Y_AG2}
Y\approx\left\{\matrix{
\sqrt{a\epsilon_e/\epsilon_B} &  1,X^2\ll a\epsilon_e/\epsilon_B \cr & \cr
a\epsilon_e/\epsilon_B & X,\,a\epsilon_e/\epsilon_B\ll 1 \cr & \cr
a\epsilon_e/(\epsilon_BX) & a\epsilon_e/\epsilon_B,\,1\ll X^2}\right. \ .
\end{equation}
For our choice of parameters, $\epsilon_B\ll\epsilon_e$ and $a\epsilon_e/\epsilon_B\gg 1$,
so that either the first or the third limits of Eq. (\ref{Y_AG2}) are relevant.
As the first limit is more often applicable, we use the parameterization
$(1+Y+X)\equiv\bar{X}(a\epsilon_e/\epsilon_B)^{1/2}$, so that the numerical 
coefficients and explicit dependence on the parameters of the break frequencies
(that depend on the electron cooling time), would be relevant for 
$X^2\ll a\epsilon_e/\epsilon_B$, where $\bar{X}\approx 1$. In the limit 
$X^2\gg a\epsilon_e/\epsilon_B$, the numerical coefficients 
and parameter dependences change because of the dependence on 
$\bar{X}\approx X/(a\epsilon_e/\epsilon_B)^{1/2}$, which is no longer 
close to 1 in this limit.

\subsection{The Synchrotron Emission}

In the standard case of a uniform ambient medium, one can express the break 
frequencies and the peak flux in terms of the shock energy $E$, the ambient 
density $n_{\rm H}$, the observed time $t$, as well as $\epsilon_e$, $\epsilon_B$, 
and the distance to the source (Sari, Piran \& Narayan 1998). This is thanks to the 
fact that for a 'standard' external medium that is composed of equal numbers of protons 
and electrons, so that both the shock dynamics (that is determined by $w=n_{\rm H}m_p c^2$) 
and the external number density of electrons (that enter the expressions for the 
flux normalization and self absorption frequency), are determined by a single parameter,
$n_{\rm H}$. In the case of a shock propagating inside a PWB, the dynamics of the AG shock 
are determined by $w=n_{\rm H, equiv}m_pc^2\approx\eta\gamma_w n_p m_pc^2$ that is dominated 
by the internal energy of the hot protons, while the number density of electrons is 
different, and dominated by the electron-positron pairs. We find
\begin{eqnarray}
n_{\rm H, equiv}&=&{\eta E_{\rm rot}\over\pi R_b^3 m_p c^2}\approx 
1.8\,\eta_{2/3}E_{\rm rot,53}\beta_{b,-1}^{-3}t_{\rm sd,1.5}^{-3}\ ,
\nonumber \\
{n_e\over n_{\rm H, equiv}}&=&{3\xi_e m_p\over 4\eta_e\gamma_w m_e}\approx
0.022\,\eta_{2/3}^{-1}\xi_{e,1/3}\gamma_{w,4.5}^{-1}\ .
\end{eqnarray}
For an elongated PWB we can make the usual substitution 
$\beta_b=3R_{\rm eff}/2ct_{\rm sd}$, to obtain the relevant expressions 
(see discussion just before \S \ref{plerion_syn}).

The self absorption frequency is typically $\nu_{sa}<\min(\nu_c,\nu_m)$, and is 
calculated using equation (\ref{alpha1}) 
for fast cooling and equation (\ref{alpha4}) for slow cooling, solving for 
$\tau_\nu'=\alpha'R_{AG}/\Gamma_{AG}=1$ for $\nu'_{sa}$ and then 
$\nu_{sa}=\Gamma_{AG}\nu'_{sa}$. The transition time from fast to slow cooling, 
$t_0$, is obtained by equating $\gamma_c$ and $\gamma_m$. For $k=0$ we get
\begin{eqnarray}
\nu_{c}&\approx & 2.1\times 10^{14}\,(1+z)^{-1/2}a^{-1}\bar{X}^{-2}
\eta_{2/3}^{-1}\epsilon_{e,-1}^{-1}\epsilon_{B,-3}^{-1/2}
E_{\rm rot,53}^{-1}\beta_{b,-1}^{3}t_{\rm sd,1.5}^{3}
E_{\rm iso,53}^{-1/2}t_{\rm days}^{-1/2}\;{\rm Hz}\ ,
\nonumber \\
\nu_{m}&\approx & 1.2\times 10^{15}\,(1+z)^{1/2}\eta_{2/3}^{2}\xi_{e,1/3}^{-2}\epsilon_{e,-1}^2
\epsilon_{B,-3}^{1/2} \gamma_{w,4.5}^2
E_{\rm iso,53}^{1/2}t_{\rm days}^{-3/2}\;{\rm Hz}\ ,
\nonumber \\
\nu_{M}&\approx & 5.2\times 10^{21}\,(1+z)^{-5/8}a^{-1/2}\bar{X}^{-1}
\eta_{2/3}^{-1/8}\epsilon_{e,-1}^{-1/2}\epsilon_{B,-3}^{1/2}
E_{\rm rot,53}^{-1/8}\beta_{b,-1}^{3/8}t_{\rm sd,1.5}^{3/8}
E_{\rm iso,53}^{1/8}t_{\rm days}^{-3/8} \;{\rm Hz}\ ,
\nonumber\\ 
\nu_{sa1}&\approx& 1.0\times 10^8 \,(1+z)^{-1/2}\bar{X}
\eta_{2/3}^{1/2}\xi_{e,1/3}^{3/5}\epsilon_{e,-1}^{1/2}\epsilon_{B,-3}^{7/10}
E_{\rm rot,53}^{11/10}\gamma_{w,4.5}^{-3/5}\beta_{b,-1}^{-3.3}t_{\rm sd,1.5}^{-3.3}
E_{\rm iso,53}^{7/10}t_{\rm days}^{-1/2}\;{\rm Hz}\ ,
 \nonumber \\
\nu_{sa2}&\approx&  2.6\times 10^7 \,(1+z)^{-1}
\eta_{2/3}^{-1}\xi_{e,1/3}^{8/5}\epsilon_{e,-1}^{-1}\epsilon_{B,-3}^{1/5}
E_{\rm rot,53}^{3/5}\gamma_{w,4.5}^{-8/5}\beta_{b,-1}^{-9/5}t_{\rm sd,1.5}^{-9/5}
E_{\rm iso,53}^{1/5}\;{\rm Hz}\ ,
 \\
F_{\nu,{\rm max}}&\approx& 0.61\,(1+z)\eta_{2/3}^{-1/2}\xi_{e,1/3}
\epsilon_{B,-3}^{1/2}
E_{\rm rot,53}^{1/2}\gamma_{w,4.5}^{-1}\beta_{b,-1}^{-3/2}t_{\rm sd,1.5}^{-3/2}
E_{\rm iso,53}d_{L28}^{-2} {\rm mJy}\ ,
\nonumber \\ \label{t_0_k0}
t_0 &\approx& 5.7 \, (1+z)\bar{X}^2\eta_{2/3}^{3}\xi_{e,1/3}^{-2}
\epsilon_{e,-1}^3\epsilon_{B,-3}E_{\rm rot,53}\gamma_{w,4.5}^2
\beta_{b,-1}^{-3}t_{\rm sd,1.5}^{-3} E_{\rm iso,53} \;{\rm days}\ ,
\nonumber \end{eqnarray}
where $\nu_{sa1}$ is for fast cooling and $\nu_{sa2}$ is for slow cooling.
For k=1 we get 
\begin{eqnarray}
\nu_{c}&\approx & 7.0\times 10^{13}\,(1+z)^{-5/6} a^{-1}\bar{X}^{-2}
\eta_{2/3}^{-4/3}\epsilon_{e,-1}^{-1}\epsilon_{B,-3}^{-1/2}
E_{\rm rot,53}^{-4/3}\beta_{b,-1}^{8/3}t_{\rm sd,1.5}^{8/3}
E_{\rm iso,53}^{-1/6}t_{\rm days}^{-1/6}\;{\rm Hz}\ ,
\nonumber \\
\nu_{m}&\approx & 9.9\times 10^{14}\,(1+z)^{1/2}\eta_{2/3}^{2}\xi_{e,1/3}^{-2}
\epsilon_{e,-1}^2\epsilon_{B,-3}^{1/2}\gamma_{w,4.5}^{2}
E_{\rm iso,53}^{1/2}t_{\rm days}^{-3/2}\;{\rm Hz}\ ,
\nonumber \\
\nu_{M}&\approx & 4.2\times 10^{21}\,(1+z)^{-2/3}a^{-1/2}\bar{X}^{-1}
\eta_{2/3}^{-1/6}\epsilon_{e,-1}^{-1/2}\epsilon_{B,-3}^{1/2}
E_{\rm rot,53}^{-1/6}\beta_{b,-1}^{1/3}t_{\rm sd,1.5}^{1/3}
E_{\rm iso,53}^{1/6}t_{\rm  days}^{-1/3} \;{\rm Hz}\ ,
\nonumber \\ 
\nu_{sa1}&\approx&3.3\times 10^8\, (1+z)^{-2/15}\bar{X}
\eta_{2/3}^{13/15}\xi_{e,1/3}^{3/5}\epsilon_{e,-1}^{1/2}\epsilon_{B,-3}^{7/10}
E_{\rm rot,53}^{22/15}\gamma_{w,4.5}^{-3/5}\beta_{b,-1}^{-44/15}t_{\rm sd,1.5}^{-44/15}
E_{\rm iso,53}^{1/3}t_{\rm days}^{-13/15}\;{\rm Hz}\ ,
\nonumber \\
\nu_{sa2}&\approx& 5.2\times 10^7 \,(1+z)^{-4/5}
\eta_{2/3}^{-4/5}\xi_{e,1/3}^{8/5}\epsilon_{e,-1}^{-1}\epsilon_{B,-3}^{1/5}
E_{\rm rot,53}^{4/5}\gamma_{w,4.5}^{-8/5}\beta_{b,-1}^{-8/5}t_{\rm sd,1.5}^{-8/5}
t_{\rm days}^{-1/5}\;{\rm Hz}\ ,
\\
F_{\nu,{\rm max}}&\approx& 1.2\,(1+z)^{7/6}\eta_{2/3}^{-1/3}\xi_{e,1/3}
\epsilon_{B,-3}^{1/2}
E^{2/3}_{\rm rot,53}\gamma_{w,4.5}^{-1}\beta_{b,-1}^{-4/3}t_{\rm sd,1.5}^{-4/3}
E_{\rm iso,53}^{5/6}t_{\rm days}^{-1/6}d_{L28}^{-2}\; {\rm mJy}\ ,
\nonumber \\ \label{t_0_k2}
t_0 &\approx& 7.3\,(1+z)\bar{X}^{3/2}\eta_{2/3}^{5/2}\xi_{e,1/3}^{-3/2}\epsilon_{e,-1}^{9/4}
\epsilon_{B,-3}^{3/4}
E_{\rm rot,53}\gamma_{w,4.5}^{3/2}\beta_{b,-1}^{-2}
t_{\rm sd,1.5}^{-2} E_{\rm iso,53}^{1/2}\;{\rm days}\ ,
\nonumber
\end{eqnarray}
We note that in order for $\nu_{sa}$ not to exceed a few GHz, as typically 
implied by observations, we need $t_{\rm sd}\gtrsim 10\;{\rm yr}$. 
This also gives more reasonable values for the transition time from fast to slow cooling, 
$t_0$, and for $F_{\nu,{\rm max}}$. For $k=0$ the effective mass density of the PWB,
becomes similar to that of a typical ISM, $n_{\rm H, equiv}=1\;{\rm cm^{-3}}$, for 
\begin{equation}\label{t_ISM}
t_{\rm ISM}=38\,\eta_{2/3}^{1/3}E_{\rm rot,53}^{1/3}\beta_{b,-1}^{-1}\;{\rm yr}\ ,
\end{equation}
while the electron number density reaches the same value for a smaller $t_{\rm sd}=
10.7\,\eta_{2/3}^{1/6}\xi_{e,1/3}^{1/6}\gamma_{w,4.5}^{-1/6}\beta_{b,-1}^{-1}\;{\rm yr}$.
For $t_{\rm sd}\sim t_{\rm ISM}$ the afterglow emission is close to that of the 'standard'
model, where the external medium is the ISM or a stellar wind, which has been extensively 
and successfully fitted to afterglow observations. 

In order to explain the X-ray lines, we need $t_{\rm sd}\lesssim t_{\rm Fe}\sim 1\;{\rm yr}$.
This implies that the radio will typically be below the self absorption 
frequency, and hence the radio emission from the afterglow would not be detectable.
On top of this, the jet break time is given by substituting $n_{\rm H, equiv}$
in place of the external density for a 'standard' external medium (Sari, Piran \& Halpern 1999),
\begin{equation}\label{t_j}
t_j=1.1\,\left({1+z\over 2}\right)\left(E_{iso,53}\over n_{H,equiv,0}\right)^{1/3}
\theta_{j,-1}^{8/3}\;{\rm days}
\end{equation}
$$
\quad\quad\quad\quad\quad\ 
=0.70\,\left({1+z\over 2}\right)
\left({E_{\rm iso,53}\over\eta_{2/3}E_{\rm rot,53}}\right)^{1/3}
\beta_{b,-1}t_{\rm sd,0}\theta_{j,-1}^{8/3}\;{\rm hr}\ ,
$$
and is very low for $\theta_{j,-1}=\theta_{j}/0.1\sim 1$. If we want to explain the observed
values of $t_j\sim 1\;{\rm day}$ that are typically observed as resulting from a larger
$\theta_j$ ($\sim 0.4$) then this would imply that the time of transition to a non-relativistic 
flow should be $t_{\rm NR}\sim\theta_j^{-2}t_j\sim 7\,t_j\sim 7\,{\rm days}$,
and in general,
\begin{equation}\label{t_NR}
t_{\rm NR}\sim{1\over c}\left({E\over n_{H,equiv}m_pc^2}\right)^{1/3}\approx
18\,\left({1+z\over 2}\right)\left({E_{51}\over \eta_{2/3}E_{\rm rot,53}}\right)^{1/3}
\beta_{b,-1}t_{\rm sd,0}\;{\rm days}\ ,
\end{equation}
where $E=10^{51}E_{51}\;{\rm ergs}$ is the true energy of the afterglow, and we have dropped
factors of order unity. Finally, for $t_{\rm sd}\lesssim t_{\rm Fe}\sim 1\;{\rm yr}$,
the transition time from fast to slow cooling is very large, and fast cooling is expected 
during all the afterglow.

For $t<t_0$, in the fast cooling regime,
the synchrotron flux density, $F_{\nu}$, is given by\footnote{If there is no
significant mixing of the shocked fluid the the spectral slope just below $\nu_{sa}$
should be $\nu^{11/8}$, and the familiar $\nu^2$ slope is obtained below a lower 
break frequency, $\nu_{ac}$ \citep{GPS00}.}:
\begin{equation}\label{FAG_syc_FC}
F_{\nu}=F_{\nu,{\rm max}}\times\left\{\matrix{
(\nu_{sa}/\nu_{c})^{1/3}(\nu/\nu_{sa})^2 & \ \ \nu<\nu_{sa} \cr & \cr 
(\nu/\nu_{c})^{1/3} & \ \ \nu_{sa}<\nu<\nu_{c} \cr & \cr 
(\nu/\nu_{c})^{-1/2} & \nu_{c}<\nu<\nu_{m} \cr & \cr 
(\nu_{m}/\nu_{c})^{-1/2}(\nu/\nu_{m})^{-p/2} & \nu_{m}<\nu<\nu_{M}}\right. \ ,
\end{equation}
For $t>t_0$ we are in the slow cooling regime, in this case  the 
spectrum peaks at $\nu_{m}$ and again consists of four power law segments:
\begin{equation}\label{FAG_syc_SC}
F_{\nu}=F_{\nu,{\rm max}}\times\left\{\matrix{
(\nu_{sa}/\nu_{m})^{1/3}(\nu/\nu_{sa})^2 & \ \ \nu<\nu_{sa} \cr & \cr 
(\nu/\nu_{m})^{1/3} & \ \ \nu_{sa}<\nu<\nu_{m} \cr & \cr 
(\nu/\nu_{m})^{(1-p)/2} & \nu_{m}<\nu<\nu_{c} \cr & \cr 
(\nu_{c}/\nu_{m})^{(1-p)/2}(\nu/\nu_{c})^{-p/2} & \nu_{c}<\nu<\nu_{M}}\right.
 \ .
\end{equation}

\subsection{The SSC Emission}

The SSC emission is calculated similarly to \S \ref{plerion_SSC} and 
\S \ref{plerion_KN}. The fast cooling spectrum is given by
\begin{equation}\label{Fnu_SSC_AG_FC}
{\nu F_{\nu}^{SC}\over\nu_mF_{\nu_m}}=Y\times\left\{\matrix{
(\nu^{SC}_{c}/\nu^{SC}_{m})^{1/2}(\nu^{SC}_{sa}/\nu^{SC}_{c})^{4/3}(\nu/\nu^{SC}_{sa})^2 & 
\ \ \nu<\nu^{SC}_{sa} \cr & \cr 
(\nu^{SC}_{c}/\nu^{SC}_{m})^{1/2}(\nu/\nu^{SC}_{c})^{4/3} & \nu^{SC}_{sa}<\nu<\nu^{SC}_{c} \cr & \cr 
(\nu/\nu^{SC}_{m})^{1/2} & \nu^{SC}_{c}<\nu<\nu^{SC}_{m} \cr & \cr 
(\nu/\nu^{SC}_{m})^{1-p/2} & \nu^{SC}_{m}<\nu<\nu_{KN,1}^{SC}\cr & \cr
(\nu_{KN,1}^{SC}/\nu^{SC}_{m})^{1-p/2}(\nu/\nu_{KN,1}^{SC})^{1/2-p}
& \nu_{KN,1}^{SC}<\nu<\nu_{KN}^{SC}(\gamma_{M})}
\right. \ ,
\end{equation}
where 
\begin{equation}\label{nuFnu_m_AG_FC}
\nu_mF_{\nu_m}=3.1\times 10^{-12}\,(1+z)\bar{X}^{-1}\epsilon_{e,-1}^{1/2}\epsilon_{B,-3}^{1/2}
  E_{\rm iso,53} t_{\rm days}^{-1} d_{L28}^{-2}
{\rm erg\,\,cm^{-2}\,s^{-1}}\ ,
\end{equation}
and for $k=0$ we have
\begin{eqnarray}
\nu^{SC}_{sa}&\approx & \gamma_{c}^2\nu_{sa1} \approx 
 {3.9\times 10^{15}\over (1+z)^{3/4}}\,\bar{X}^{-1}
\eta_{2/3}^{-3/4}\xi_{e,1/3}^{3/5}\epsilon_{e,-1}^{-1/2}
\epsilon_{B,-3}^{-3/10}
E_{\rm rot,53}^{-3/20}\gamma_{w,4.5}^{-3/5}\beta_{b,-1}^{9/20}t_{\rm sd,1.5}^{9/20}
E_{\rm iso,53}^{-1/20}t_{\rm days}^{-1/4}\;{\rm Hz}\ ,
\nonumber \\
\nu^{SC}_{c}& \approx & \gamma_{c}^2\nu_{c}\approx
  {1.5\times 10^{22} \over (1+z)^{3/4}}\,a^{-2}\bar{X}^{-4}
\eta_{2/3}^{-9/4}\epsilon_{e,-1}^{-2}\epsilon_{B,-3}^{-3/2}
E_{\rm rot,53}^{-9/4}\beta_{b,-1}^{27/4}t_{\rm sd,1.5}^{27/4}
E_{\rm iso,53}^{-5/4}t_{\rm days}^{-1/4}\; {\rm Hz}\ ,
\nonumber \\ \label{freq_AG_FC_k0}
\nu^{SC}_{m}& \approx & \gamma_{m}^2\nu_{m}\approx  
{5.1\times 10^{23}\over(1+z)^{-5/4}}\,\eta_{2/3}^{15/4}\xi_{e,1/3}^{-4}
\epsilon_{e,-1}^{4}\epsilon_{B,-3}^{1/2}
 E_{\rm rot,53}^{-1/4}\gamma_{w,4.5}^{4}\beta_{b,-1}^{3/4}t_{\rm sd,1.5}^{3/4}E_{\rm iso,53}^{3/4}
t_{\rm days}^{-9/4}
\;{\rm Hz}\
\\
\nu_{KN,1}^{SC}&\approx&{\Gamma_{AG}^2\nu_{KN,0}^2\over\nu_m} \approx \;
 {6.7\times 10^{26}\over(1+z)^{7/4}}\,\eta_{2/3}^{-9/4}\xi_{e,1/3}^{2} 
\epsilon_{e,-1}^{-2}\epsilon_{B,-3}^{-1/2}
E_{\rm rot,53}^{-1/4}\gamma_{w,4.5}^{-2}\beta_{b,-1}^{3/4}t_{\rm sd,1.5}^{3/4} 
E_{\rm iso,53}^{-1/4}t_{\rm days}^{3/4}\;{\rm Hz}\ ,
\nonumber \\
\nu_{KN}^{SC}(\gamma_{M})&\approx&\nu_{KN,0}\Gamma_{AG}\gamma_{M}\approx 
{3.8\times 10^{28}\over(1+z)^{13/16}}\,a^{-1/4}\bar{X}^{-1/2}\eta_{2/3}^{-5/16}
\epsilon_{e,-1}^{-1/4}E_{\rm rot,53}^{-5/16}\beta_{b,-1}^{15/16}
t_{\rm sd,1.5}^{15/16}E_{\rm iso,53}^{1/16}t_{\rm days}^{-3/16}\;{\rm Hz}\ ,
\nonumber
\end{eqnarray}
while for $k=1$ we have
\begin{eqnarray}
\nu^{SC}_{sa}&\approx& {3.3\times 10^{15}\over(1+z)^{4/5}} \,\bar{X}^{-1}
\eta_{2/3}^{-4/5}\xi_{e,1/3}^{3/5}\epsilon_{e,-1}^{-1/2}\epsilon_{B,-3}^{-3/10}
E_{\rm rot,53}^{-1/5}\gamma_{w,4.5}^{-3/5}\beta_{b,-1}^{2/5}t_{\rm sd,1.5}^{2/5}
t_{\rm days}^{-1/5}
\;{\rm Hz}\ ,
\nonumber \\
\nu^{SC}_{c}& \approx &
 {1.3\times 10^{21} \over(1+z)^{3/2}}\,a^{-2}\bar{X}^{-4}
\eta_{2/3}^{-3}\epsilon_{e,-1}^{-2}\epsilon_{B,-3}^{-3/2}
E_{\rm rot,53}^{-3}\beta_{b,-1}^{6}t_{\rm sd,1.5}^{6}
E_{\rm iso,53}^{-1/2}t_{\rm days}^{1/2}\; {\rm Hz}\ ,
\nonumber \\ \label{freq_AG_FC_k1}
\nu^{SC}_{m}& \approx &   
{2.7\times 10^{23}\over(1+z)^{-7/6}}\,\eta_{2/3}^{11/3}\xi_{e,1/3}^{-4}\epsilon_{e,-1}^4 
\epsilon_{B,-3}^{1/2}
E_{\rm rot,53}^{-1/3}\gamma_{w,4.5}^{4}\beta_{b,-1}^{2/3}t_{\rm sd,1.5}^{2/3}E_{\rm iso,53}^{5/6}
t_{\rm days}^{-13/6}
\;{\rm Hz}\
\\
\nu_{KN,1}^{SC}&\approx&
{5.3\times  10^{26}\over(1+z)^{11/6}}\,\eta_{2/3}^{-7/3}\xi_{e,1/3}^{2}\epsilon_{e,-1}^{-2}
\epsilon_{B,-3}^{-1/2} E_{\rm rot,53}^{-1/3}\gamma_{w,4.5}^{-2}
\beta_{b,-1}^{2/3}t_{\rm sd,1.5}^{2/3}E_{\rm iso,53}^{-1/6}
t_{\rm days}^{5/6}\;{\rm Hz}\ ,
\nonumber \\
\nu_{KN}^{SC}(\gamma_{M})&\approx& 
{2.5\times 10^{28}\over(1+z)^{11/12}}\,a^{-1/4}\bar{X}^{-1/2}\eta_{2/3}^{-5/12}\epsilon_{e,-1}^{-1/4} 
E_{\rm rot,53}^{-5/12}\beta_{b,-1}^{5/6}t_{\rm sd,1.5}^{5/6}
E_{\rm iso,53}^{1/6}t_{\rm days}^{-1/12}
\;{\rm Hz}\ .
\nonumber
\end{eqnarray}
The slow cooling spectrum is
\begin{equation}\label{Fnu_SSC_AG_SC}
{\nu F_{\nu}^{SC}\over\nu_cF_{\nu_c}}=Y\times\left\{\matrix{
(\nu^{SC}_{m}/\nu^{SC}_{c})^{(3-p)/2}(\nu^{SC}_{sa}/\nu^{SC}_{m})^{4/3}(\nu/\nu^{SC}_{sa})^2 & 
\ \ \nu<\nu^{SC}_{sa} \cr & \cr 
(\nu^{SC}_{m}/\nu^{SC}_{c})^{(3-p)/2}(\nu/\nu^{SC}_{m})^{4/3} & \nu^{SC}_{sa}<\nu<\nu^{SC}_{m} \cr & \cr 
(\nu/\nu^{SC}_{c})^{(3-p)/2} & \nu^{SC}_{m}<\nu<\nu^{SC}_{c} \cr & \cr 
(\nu/\nu^{SC}_{c})^{1-p/2} & \nu^{SC}_{c}<\nu<\nu_{KN,1}^{SC}\cr & \cr
(\nu_{KN,1}^{SC}/\nu^{SC}_{c})^{1-p/2}(\nu/\nu_{KN,1}^{SC})^{-(p+1)/2}
& \nu_{KN,1}^{SC}<\nu<\nu_{KN}^{SC}(\gamma_{M})}
\right. \ .
\end{equation}
where $\nu_cF_{\nu_c}$ is just $a=(\nu_m/\nu_c)^{(2-p)/2}$ times $\nu_mF_{\nu_m}$ for 
the fast cooling, that is given in Eq. (\ref{nuFnu_m_AG_FC}). For $k=0$ we have
\begin{eqnarray}
\nu^{SC}_{sa}&\approx & \gamma_{m}^2\nu_{sa2} \approx 
 {1.1\times 10^{16}\over(1+z)^{1/4}}\,\eta_{2/3}^{3/4}\xi_{e,1/3}^{-2/5}
\epsilon_{e,-1}\epsilon_{B,-3}^{1/5}
E_{\rm rot,53}^{7/20}\gamma_{w,4.5}^{2/5}\beta_{b,-1}^{-21/20}
t_{\rm sd,1.5}^{-21/20}E_{\rm iso,53}^{9/20}t_{\rm days}^{-3/4}  
\;{\rm Hz}\ ,\nonumber
\\
\nu_{KN,1}^{SC}&=&{\Gamma_{AG}^2\nu_{KN,0}^2\over\nu_c} \approx 
{3.8\times 10^{27}\over(1+z)^{3/4}} \,a \bar{X}^{2}\eta_{2/3}^{3/4}\epsilon_{e,-1}\epsilon_{B,-3}^{1/2}
E_{\rm rot,53}^{3/4}\beta_{b,-1}^{-9/4}t_{\rm sd,1.5}^{-9/4}
E_{\rm iso,53}^{3/4}t_{\rm days}^{-1/4}
\;{\rm Hz}\ ,
\end{eqnarray}
and for $k=1$ we obtain
\begin{eqnarray}
\nu^{SC}_{sa}&\approx &  
 {1.4\times 10^{16}\over(1+z)^{2/15}}\,\eta_{2/3}^{13/15}\xi_{e,1/3}^{-2/5}\epsilon_{e,-1}
\epsilon_{B,-3}^{1/5}
E_{\rm rot,53}^{7/15}\gamma_{w,4.5}^{2/5}\beta_{b,-1}^{-14/15}
t_{\rm sd,1.5}^{-14/15}E_{\rm iso,53}^{1/3}t_{\rm days}^{-13/15}  
\;{\rm Hz}\ ,
\nonumber \\
\nu_{KN,1}^{SC}&\approx& 
{7.6\times 10^{27}\over(1+z)^{1/2}}\,a\bar{X}^{2}\eta_{2/3}\epsilon_{e,-1}
\epsilon_{B,-3}^{1/2}E_{\rm rot,53}\beta_{b,-1}^{-2}t_{\rm sd,1.5}^{-2}
E_{\rm iso,53}^{1/2}t_{\rm days}^{-1/2}
\;{\rm Hz}\ ,
\end{eqnarray}

\subsection{The EC Emission}
\label{AG_EC}

The EC emission in this case arises from the upscattering of the plerion radiation
by the relativistic electrons behind the afterglow shock. We provide detailed 
expressions for one representative plerion spectrum, the one given in equation 
(\ref{Fnu_syc_FC}). This spectrum is the spectrum for 
$12\:{\rm yr}\sim t_2<t_{\rm sd}<t_3\sim 65\:{\rm yr}$ (see equations \ref{t_2}, \ref{t_3}), 
which is of most interest. Similar expressions for the 
other plerion spectra can be readily derived in a similar manner. We note that for 
$t_{\rm sd}<t_2$ the synchrotron emission of the plerion near the peak of $\nu F_\nu$ is
the same as for the spectrum we use (i.e. for $t_2<t_{\rm sd}<t_3$), and therefore the EC 
near the peak of its $\nu F_\nu$ should be the same.  The peak of $\nu F_\nu$ for the SSC
plerion emission is typically above the KN limit for the AG electrons, and should therefore
have a negligible contribution for the EC emission of the afterglow.
The resulting EC spectrum depends on whether the electrons in the afterglow shock are in 
the fast cooling or slow cooling regime. The EC spectrum is
\begin{equation}\label{Fnu_EC_AG_FC}
{\nu F_{\nu}^{EC}\over a\,\nu_m F_{\nu_m}}=X\times\left\{\matrix{
(\nu_{c}^{EC}/\nu^{EC}_{m})^{1/2}(\nu_{sa}^{EC}/\nu_{c}^{EC})^{4/3}(\nu/\nu_{sa}^{EC})^2 & 
\nu<\nu_{sa}^{EC} \cr & \cr
(\nu_{c}^{EC}/\nu^{EC}_{m})^{1/2}(\nu/\nu_{c}^{EC})^{4/3} & 
\nu_{sa}^{EC}<\nu<\nu_{c}^{EC} \cr & \cr 
(\nu/\nu^{EC}_{m})^{1/2} & \nu_{c}^{EC}<\nu<\nu^{EC}_{m}\cr & \cr
(\nu/\nu^{EC}_{m})^{1-s/2} & \nu^{EC}_{m}<\nu<\nu_{KN}^{EC}(\gamma_{M})}
\right. \ .
\end{equation}
where $\nu_{KN}^{EC}(\gamma_{M})=\nu_{KN}^{SC}(\gamma_{M})$ is given 
by equation (\ref{freq_AG_FC_k0}). For fast cooling  and  k=0
\begin{eqnarray}
\nu_{sa}^{EC}&\approx & \Gamma_{AG}^2\gamma_{c}^2\nu_{bsa} \approx 
{2.4\times10^{18}\over\sqrt{1+z}}{\xi_{e,1/3}^{1/3}\epsilon_{bB,-3}^{1/6}
\beta_{b,-1}^{23/6}t_{\rm sd,1.5}^{7/2}\over
\bar{X}^2\eta_{2/3}^{3/2}\epsilon_{be,1/3}^{1/6}\epsilon_{e,-1}
\epsilon_{B,-3}E_{\rm rot,53}^{7/6}\gamma_{w,4.5}^{1/3}E_{\rm iso,53}^{1/2}t_{\rm days}^{1/2}}
\;{\rm Hz}\ ,
\nonumber \\ \label{freq_EC_AG_FC_k0}
\nu_{c}^{EC}&\approx & \Gamma_{AG}^2\gamma_{c}^2\nu_{bc} \approx 
{6.4\times10^{19}\over\sqrt{1+z}}\,{\beta_{b,-1}^{9}t_{\rm sd,1.5}^{7}\over
\bar{X}^2\eta_{2/3}^{3}\epsilon_{be,1/3}\epsilon_{bB,-3}^{1/2}\epsilon_{e,-1}\epsilon_{B,-3}
E_{\rm rot,53}^{3}E_{\rm iso,53}^{1/2}t_{\rm days}^{1/2}} 
\;{\rm Hz}\ ,
\\
\nu^{EC}_{m}& \approx & \Gamma_{AG}^2\gamma_{m}^2\nu_{bm}\approx  
{6.5\times 10^{21}\over(1+z)^{-1/2}}\,\eta_{2/3}^{4}\xi_{e,1/3}^{-4}
\epsilon_{be,1/3}^{2}\epsilon_{bB,-3}^{1/2}\epsilon_{e,-1}^{2}
\gamma_{w,4.5}^{4}E_{\rm iso,53}^{1/2}t_{\rm days}^{-3/2} 
\;{\rm Hz}\ ,\nonumber
\end{eqnarray}
while for fast cooling with $k=1$ we have
\begin{eqnarray}
\nu_{sa}^{EC}&\approx & 
{4.1\times 10^{17}\over(1+z)}\,\bar{X}^{-2}\eta_{2/3}^{-2}\xi_{e,1/3}^{1/3}\epsilon_{be,1/3}^{-1/6}
\epsilon_{bB,-3}^{1/6}\epsilon_{e,-1}^{-1}\epsilon_{B,-3}^{-1}E_{\rm rot,53}^{-5/3}
\gamma_{w,4.5}^{-1/3}\beta_{b,-1}^{10/3}t_{\rm sd,1.5}^{3} 
\;{\rm Hz}\ , \nonumber
\nonumber \\
\nu_{c}^{EC}&\approx & 
{1.1\times 10^{19}\over(1+z)}\,\bar{X}^{-2}\eta_{2/3}^{-7/2}\epsilon_{be,1/3}^{-1}
\epsilon_{bB,-3}^{-1/2}\epsilon_{e,-1}^{-1}\epsilon_{B,-3}^{-1}
E_{\rm rot,53}^{-7/2}\beta_{b,-1}^{17/2}t_{\rm sd,1.5}^{13/2}
\;{\rm Hz}\ ,
\\
\nu^{EC}_{m}& \approx &  
{2.7\times10^{21}\over(1+z)^{-1/3}}\,\eta_{2/3}^{23/6}\xi_{e,1/3}^{-4}\epsilon_{be,1/3}^{2}
\epsilon_{bB,-3}^{1/2}\epsilon_{e,-1}^{2}E_{\rm rot,53}^{-1/6}\gamma_{w,4.5}^{4}
\beta_{b,-1}^{-5/6}t_{\rm sd,1.5}^{-5/6}E_{\rm iso,53}^{2/3}t_{\rm days}^{-4/3} 
\;{\rm Hz}\ .\nonumber
\end{eqnarray}
For slow cooling with $k=0$ we have
 \begin{eqnarray}
\nu_{sa}^{EC}&\approx & \Gamma_{AG}^2\gamma_{m}^2\nu_{bsa} \approx 
{1.4\times 10^{19}\over(1+z)^{-1/2}}{\eta_{2/3}^{3/2}\epsilon_{bB,-3}^{1/6}
\epsilon_{e,-1}^{2}\gamma_{w,4.5}^{5/3}\beta_{b,-1}^{5/6}
t_{\rm sd,1.5}^{1/2}E_{\rm iso,53}^{1/2} \over
\xi_{e,1/3}^{5/3}\epsilon_{be,1/3}^{1/6}E_{\rm rot,53}^{1/6}t_{\rm days}^{3/2}}
\;{\rm Hz}\ ,\nonumber
\\
\nu_{c}^{EC}&\approx & \Gamma_{AG}^2\gamma_{m}^2\nu_{bc} \approx 
{3.7\times 10^{20}\over(1+z)^{-1/2}}{\epsilon_{e,-1}^{2}\gamma_{w,4.5}^{2}
\beta_{b,-1}^{6}t_{\rm sd,1.5}^{4}E_{\rm iso,53}^{1/2} \over
\xi_{e,1/3}^{2}\epsilon_{be,1/3}\epsilon_{bB,-3}^{1/2}E_{\rm rot,53}^{2}t_{\rm days}^{3/2}}
\;{\rm Hz}\ ,
\\
\nu^{EC}_{m}& \approx & \Gamma_{AG}^2\gamma_{c}^2\nu_{bm}\approx  
{1.1\times 10^{21}\over\sqrt{1+z}}{\eta_{2/3}\epsilon_{be,1/3}^{2}
\epsilon_{bB,-3}^{1/2}\gamma_{w,4.5}^{2}\beta_{b,-1}^{3}t_{\rm sd,1.5}^{3}
\over a\,\bar{X}^{2}\xi_{e,1/3}^{2}\epsilon_{e,-1}\epsilon_{B,-3}E_{\rm rot,53}
E_{\rm iso,53}^{1/2}t_{\rm days}^{1/2}} 
\;{\rm Hz}\ ,\nonumber
\end{eqnarray}
and finally for $k=1$ we have
\begin{eqnarray}
\nu_{sa}^{EC}&\approx & 
{5.8\times 10^{18}\over(1+z)^{-1/3}}\,\eta_{2/3}^{4/3}\xi_{e,1/3}^{-5/3}
\epsilon_{be,1/3}^{-1/6}\epsilon_{bB,-3}^{1/6}\epsilon_{e,-1}^{2}
E_{\rm rot,53}^{-1/3}\gamma_{w,4.5}^{5/3}\beta_{b,-1}^{2/3}
t_{\rm sd,1.5}^{1/3}E_{\rm iso,53}^{2/3}t_{\rm days}^{-4/3}
\;{\rm Hz}\ ,
\nonumber \\
\nu_{c}^{EC}&\approx & 
{1.6\times 10^{20}\over(1+z)^{-1/3}}\,\eta_{2/3}^{-1/6}\xi_{e,1/3}^{-2}
\epsilon_{be,1/3}^{-1}\epsilon_{bB,-3}^{-1/2}\epsilon_{e,-1}^{2}
E_{\rm rot,53}^{-13/6}\gamma_{w,4.5}^{2}\beta_{b,-1}^{35/6}
t_{\rm sd,1.5}^{23/6}E_{\rm iso,53}^{2/3}t_{\rm days}^{-4/3}
\;{\rm Hz}\ ,
\\
\nu^{EC}_{m}& \approx & 
{1.9\times 10^{20}\over(1+z)}\,a^{-1}\bar{X}^{-2}\eta_{2/3}^{1/2}\xi_{e,1/3}^{-2} 
\epsilon_{be,1/3}^{2}\epsilon_{bB,-3}^{1/2}\epsilon_{e,-1}^{-1}\epsilon_{B,-3}^{-1}
E_{\rm rot,53}^{-3/2}\gamma_{w,4.5}^{2}\beta_{b,-1}^{5/2}t_{\rm sd,1.5}^{5/2}
\;{\rm Hz}\ ,\nonumber
\end{eqnarray}

\subsection{High Energy Emission}

Figure \ref{fig2} shows the $\nu F_\nu$ spectrum of the afterglow at $t=500\;{\rm s}$ 
and $5\times 10^3\;{\rm s}$, for our fiducial parameters, and for 
$t_{\rm sd}=20\;{\rm yr}$, $E_{\rm rot,53}=0.5$, $z=1$, $R_s/R_b=0.1$. 
As can be seen from the figure, for $t=500\;{\rm s}$ ($5\times 10^3\;{\rm s}$) 
the synchrotron is dominant below $45\;$MeV ($2\;$MeV), while the EC is dominant 
above this range. At $t=5\times 10^3\;{\rm s}$ the SSC component becomes dominant 
above $40\;$GeV. We expect an upper cutoff at $h\nu_{\gamma\gamma}\sim 250\;$GeV, 
due to opacity to pair production, with the photons of the plerion. 
This upper cutoff moves down to a lower energy for smaller values of $t_{\rm sd}$, 
and is $\sim 1\;$GeV for $t_{\rm sd}=1\;{\rm yr}\sim t_{\rm Fe}$. 
This implies that for afterglows with X-ray line features we expect no high 
energy emission above this limit.

We find that the early afterglow ($t\lesssim 100\;$s) emission at 
$\gtrsim 100\;$MeV is dominated by the EC and SSC component, which are 
comparable at this time. At later times the EC becomes dominant over the SSC 
component. The peak of the $\nu F_\nu^{EC}$ emission is at the 
level of $\sim 5\times 10^{-9}(t/500\;{\rm s})^{-1}\;{\rm ergs\;cm^{-2}\,s^{-1}}$, 
and is located at $h\nu_m^{EC}\sim 70\,(t/500\;{\rm s})^{-3/2}\;$GeV 
(see Eq. \ref{freq_EC_AG_FC_k0}). The spectrum scales as $\nu F_\nu\propto\nu^{(2-s)/2}$ 
above this photon energy, implying a flat $\nu F_\nu$ for values of $s\sim2$ that are 
typically inferred for PWBs, while it scales as $\nu F_\nu\propto\nu^{1/2}$ 
below this energy. At early times the afterglow radius is relatively small 
($R_{AG}\lesssim fR_b$), so that $X$ is approximately constant in time, and the 
peak of the $\nu F_\nu$ EC spectrum has a temporal
scaling similar to that for the synchrotron component (i.e. $\propto t^{-1}$, see Eq. 
\ref{nuFnu_m_AG_FC}). We therefore expect $\nu F_\nu$ at a fixed photon energy to decay 
very slowly with time, as $t^{-1/4}$, at $\nu<\nu_m^{EC}$, and decay approximately linearly 
with time ($\propto t^{-1-3(s-2)/4}$) at $\nu>\nu_m^{EC}$. The temporal decay becomes larger
than these scalings as the afterglow radius $R_{AG}$ increases above $fR_b$, and the parameter
$X$ begins to decrease with time. 

This result can explain the high energy emission detected by EGRET for GRB 940217 (Hurley et al. 1994).
This detection consists of photons of energies $\gtrsim 50\;$MeV, where 10 photons were observed
during the prompt GRB emission, that lasted $180\;$s, and 18 photons were detected up to 
$5,400\;$s after the end of the GRB, which include a photon  of energy $18\;$GeV. 
During the post-GRB emission, the source position was Earth-occulted for $\sim 3,700\;$s.
At $t\sim 500\;$s the flux is $\sim 1-2\times 10^{-9}\;{\rm ergs\;cm^{-2}\,s^{-1}}$
which is roughly consistent with our results. At $t\sim 5,000\;$s, after the Earth-occultation, 
the flux is a factor of $\sim 2-3$ lower, if we exclude the one $18\;$GeV photon. 
This moderate time decay is consistent with our results.

A different interpretation for the high energy emission discussed above was 
recently suggested by Wang, Dai \& Lu (2002), in a similar context of the 
supranova model, where the GRB occurs inside a plerionic environment. 
The main difference is that they consider a pulsar wind that consists purely 
of $e^\pm$ pairs, that is in the fast cooling regime, and therefore the radius
of the termination shock of the wind, $R_s$ is very close to the outer radius 
of the PWB, $R_b$, and all the shocked wind is concentrated within a thin radial 
interval, $R_b-R_s\ll R_b$. They try to explain the high energy emission as the 
synchrotron emission from the early afterglow. They obtain $h\nu_m\sim 1\;$GeV 
at $t\sim 180\;$s, and according to their model $t\sim R_b/2\Gamma^2c$ or 
$\Gamma\propto t^{-1/2}$ and $\nu_m\propto\Gamma^4\propto t^{-2}$. However, 
this implies $h\nu_m\sim$ a few keV after one day, which is inconsistent with 
afterglow observations. They also claim that the EC emission is unimportant, 
which is in contradiction with our conclusions. The inclusion of a proton component 
in the pulsar wind with a similar energy to that of the $e^\pm$ 
component allows only the energy in the $e^\pm$ to be radiated away, so that even 
for a fast cooling PWB a large part of the energy of the pulsar wind remains in 
the protons and $R_s$ is significantly smaller than $R_b$.

\section{Discussion}
\label{discussion}

We have studied the observational implications of GRBs occurring inside a pulsar wind 
bubble (PWB), as is expected in the supranova model. We find that the most important 
parameter that determines the behavior of the system is the time delay, $t_{\rm sd}$, 
between the supernova and GRB events. The value of $t_{\rm sd}$ is given by the typical 
timescale on which the SMNS loses its rotational energy due to magnetic dipole radiation 
(see Eq. \ref{t_sd}) and depends mainly on the magnetic field strength of the SMNS, $B_*$ 
(since its mass, radius and spin period are constrained to a much smaller range of 
possible values). For $B_*\sim 10^{12}-10^{13}\;{\rm G}$,  $t_{\rm sd}$ is between 
a few weeks and several years. However, a larger range in $B_*$, and correspondingly 
in $t_{\rm sd}$, seems plausible. We therefore consider $t_{\rm sd}$ as a free parameter, 
and predict the observational consequences of different values for this parameter:

1. For extremely small values of $t_{\rm sd}<t_{\rm col}=R_\star/\beta_b c\approx 
0.9\,R_{\star,13}\beta_{b,-1}^{-1}\;{\rm hr}$, where $R_\star=10^{13}R_{\star,13}\;{\rm cm}$
is the radius of the progenitor star (before it explodes in a supernova), the stellar 
envelope does not have enough time to increase its radius considerably before the GRB 
goes off, and the supranova model reduces to the collapsar model. In this respect, the 
collapsar model may be seen as a special case of the supranova model. Such low values 
of $t_{\rm sd}$ might be achieved if the SMNS is not rotating uniformly, as differential 
rotation may amplify the magnetic field to very large values.

2. When $t_{\rm col}<t_{\rm sd}<t_{\rm IS}\sim 3\beta_b^{-1}t_{v}\Gamma^{2(4-k)\over 
(3-k)}\approx 16\beta_{b,-1}^{-1}\Gamma_{2.5}^{8/3}t_{v,-2}\;{\rm days}$ (for $k=0$) the 
deceleration radius $R_{\rm dec}\sim R_{\rm NR}\Gamma^{-2/(3-k)}\sim R_b\Gamma^{-2/(3-k)}$ 
is smaller than the radius for internal shocks $R_{\rm IS}\sim 2\Gamma^2ct_v$. In this 
case the kinetic energy of the GRB ejecta is dissipated through an external shock that 
is driven into the shocked pulsar wind, before internal shocks that result from 
variability within the outflow have time to occur. For an elongated PWB, $t_{\rm IS}$ can
be smaller by up to a factor for $\sim 10$, since the polar radius would be $\sim 10$
time larger for the same $t_{\rm sd}$, and the volume of the PWB would be much larger, 
and the density much smaller.

3. If $t_{\rm IS}<t_{\rm sd}<t_\tau\sim 0.4\;{\rm yr}$, internal shock can occur inside 
the PWB, but the SNR shell is still optically thick to Thomson scattering, and the 
radiation from the plerion, the prompt GRB and the afterglow cannot escape and reach 
the observer. If the SNR shell is clumpy (possibly due to the Rayleigh-Taylor 
instability, see \S \ref{PWB}), then the Thomson optical  depth in the under-dense 
regions within the shell may decrease below unity at $t_{\rm sd}$ somewhat smaller 
than $t_\tau$, enabling some of the radiation from the plerion to escape. For an elongated 
PWB, the polar radius can be larger by up to a factor of $\sim 10$, which reduces $t_\tau$
by the same factor. Furthermore, the elongation can be due to a smaller than average 
surface mass density of the SNR shell at the poles. This would further reduce $t_\tau$.

4. For $t_\tau<t_{\rm sd}<t_{\rm Fe}\sim 1\;{\rm yr}$ the SNR shell has a Thomson 
optical depth smaller than unity, but the optical depth for the iron line features 
is still $\gtrsim 1$ so that detectable X-ray line features, like the iron lines 
observed in several afterglows, can be produced.

5. Finally, for $t_{\rm Fe}<t_{\rm sd}$, we expect no iron lines. When $t_{\rm sd}$ 
is between $\sim 2\;{\rm yr}$ and $\sim 20\;{\rm yr}$ the radio emission of the 
plerion may be detectable for $\gamma_w\lesssim 10^5$. The lack of detection of such 
a radio emission excludes values of  $t_{\rm sd}$ in this range, if indeed 
$\gamma_w\lesssim 10^5$, as is needed to obtain reasonable values for the break 
frequencies of the afterglow. For $t_{\rm sd}=t_{\rm ISM}\sim 38\;{\rm yr}$, \
the effective density of the PWB is similar to that of the ISM (i.e. $1\;{\rm cm^{-3}}$), 
and the afterglow emission is similar to that of the standard model, where $k=0$ is 
similar to an ISM environment, with the exception that in our model a value of $k=1$, 
that is intermediate between an ISM and a stellar wind, is also possible. 
Larger (smaller) values of the external density are obtained for smaller (larger)
values of $t_{\rm sd}$.

The SNR shell is decelerated due to the sweeping up of the surrounding medium for 
$t_{\rm sd}>225\,M_{\rm SNR,1}^{1/3}n_0^{-1/3}\beta_{b,-1}^{-1}\;{\rm yr}$, where 
$n=n_0\;{\rm cm^{-3}}$ is the number density of the external medium, which is larger 
than the values of $t_{\rm sd}$ that are of interest to us. This effect may therefore 
be neglected for our purposes. 

An important difference between our analysis and previous works (KG; IGP) is that 
we allow for a proton component in the pulsar wind, that carries a significant 
fraction of its energy. In contrast to the $e^\pm$ component, the internal 
energy of the protons in the shocked wind is not radiated away, and therefore a 
large fraction of the energy in the pulsar wind ($\sim 10^{53}\;{\rm ergs}$) 
is always left in the PWB. This implies that even for a fast cooling PWB, 
the radius of the wind termination shock $R_s$ is significantly smaller than 
the radius of the SNR, $R_b$, and that the afterglow shock typically becomes 
non-relativistic before it reaches the outer boundary of the PWB.

\section{Conclusions}
\label{conclusions}

Our main conclusion is that existing afterglow observations put interesting constraints
on the model parameters, the most important of which being the time delay $t_{\rm sd}$ between
the supernova and GRB events, which is constrained to be $\gtrsim 20\;{\rm yr}$, in order
to explain typical afterglow observations and the lack of detection of the plerion emission
in the radio during the afterglow. Another important conclusion is that iron line features that 
have been observed in a few X-ray afterglows cannot be naturally explained within the simplest 
spherical version of the PWB model, that has been considered in this work. This is because the 
production of these lines requires $t_{\rm sd}\lesssim t_{\rm Fe}\sim 1\;{\rm yr}$ which implies 
a very large density for the PWB and effects the afterglow emission in a number of different ways: 
i) The self absorption frequency of the afterglow is typically above the radio,
implying no detectable radio afterglow, while radio afterglows were detected for
GRBs 970508, 970828, and 991216, where the iron line feature for the latest of these three
is the most significant detection to date ($\sim 4\sigma$).
We also expect the self absorption frequency of the plerion emission to be above the
radio in this case, so that the radio emission from the plerion should not be detectable,
and possibly confused with that of the afterglow. 
ii) A short jet break time $t_j$ and a relatively short non-relativistic transition time
$t_{\rm NR}$ are implied, as both scale linearly with $t_{\rm sd}$ and are in the right 
range inferred from observations for $t_{\rm sd}\sim 30\;{\rm yr}$ (see Eqs. \ref{t_j}, \ref{t_NR}).
iii) The electrons are always in the fast cooling regime during the entire afterglow.

The above constraints regarding the iron lines may be relaxed if we allow for deviations from the 
simple spherical geometry we have assumed for the PWB. A natural variant is when the PWB 
becomes elongated along its rotational axis (KG). This may occur if the surface mass density of 
the SNR shell is smaller at the poles compared to the equator, so that during the acceleration of 
the SNR shell by the pressure of the shocked pulsar wind (that is expected to be roughly the same at 
the poles and at the equator) its radius will become larger at the poles, as the acceleration there 
will be larger. A large-scale toroidal magnetic field within the PWB may also contribute to the 
elongation of the SNR shell along its polar axis (KG). It is also likely that the progenitor star
that gave rise to a SMNS had an anisotropic mass loss, which results in a density contrast between 
the equators (where the density is higher) and the poles (where the density is lower). A sufficiently
large density contrast between the equator and the poles can also contribute to the elongation of the
shell, for sufficiently large $t_{\rm sd}$, as the SNR shell will begin to be decelerated due to the
interaction with the external medium, at a smaller radius near the equator, compared to the poles.
A similar non-spherical variant of the model is if we allow for holes in the SNR shell, that extend 
over a small angle around the polar axis, where all the wind is decelerated in a 
termination shock within the SNR shell ($R_s<R_b$), but most of the shocked pulsar wind can get out 
through the holes near the poles and reach a radius considerably larger than $R_b$. This variant
may be viewed as a limiting case of the previous variant, when the surface density contrast of the 
SNR shell, between the equator and the poles is very large. This implies that most of the mass in 
the SNR shell is concentrated near the equator, while only a small fraction of it is near the poles,
so that the radius near the poles can be as large as $\sim ct_{\rm sd}$, while the equatorial radius
is $\lesssim R_b$. In both variants, the total volume of the PWB is close to that of a sphere 
with the polar radius, and much larger than that of a sphere with the equatorial radius. This would 
allow for a smaller density with the same small value of the equatorial radius that is required to 
produce the iron lines. We see that in principle, variants of the simple model are capable of 
reconciling between the iron line detections and the afterglow observations.

An important advantage of the PWB model is that it can naturally explain the large values of 
$\epsilon_B$ and $\epsilon_e$ that are inferred from fits to afterglow data (KG), thanks to 
the large relative number of electron-positron pairs and large magnetic fields in the PWB. 
This is in contrast with standard environment that is usually assumed to be either an ISM or 
the stellar wind of a massive progenitor, that consists of protons and electrons, and where 
the magnetic field is too small to explain the values inferred from afterglow observations, 
and magnetic field generation at the shock itself is required.  
Additional advantages of the PWB model are its ability to naturally account for the range of
external densities inferred from afterglow fits, and allowing for a homogeneous external medium 
($k\sim 0$), as inferred for most afterglows, with a massive star progenitor.

Another advantage of the PWB model is its capability of explaining the high 
energy emission observed in some GRBs (Schneid et al. 1992; Sommer et al. 1994; 
Hurley et al. 19994; Schneid et al. 1995). We find that the high energy emission 
during the early afterglow at photon energies $\gtrsim 100\;$keV is dominated by 
the external Compton (EC) component, that is due to the upscattering of photons 
from the plerion radiation field to higher energies by the relativistic electrons 
behind the afterglow shock. We predict that such a high energy emission may be 
detected in a large fraction of GRBs with the upcoming mission GLAST. However, 
we find an upper cutoff at a photon energy of $\sim 1\,t_{\rm sd,0}^2\;$GeV,
due to opacity to pair production with the photons of the PWB. 
This implies no high energy emission above $\sim 1\;$GeV for afterglows 
with X-ray line features, but allows photons up to an energy of $\sim 1\;$TeV 
for afterglows with an external density typical of the ISM 
($t_{\rm sd}\sim t_{\rm ISM}$).

\acknowledgements

We thank Arieh K\"onigl for his careful reading of the manuscript and helpful comments.
This research was supported by the partial support of
the Italian Ministry for University and Research
(MIUR) through the grant Cofin-01-02-43 (DG) and by the grant NSF PHY 00-70928 (JG).
We thank the Einstein Center at the Weizmann Institute of Science for the 
hospitality and for the pleasant working atmosphere.
DG thanks the Institute for Advanced Study, where most of this research was carried out,
for the hospitality and the nice working atmosphere.

\newpage

\end{document}